\documentclass[twoside,twocolumn,9pt]{article}
\usepackage{extsizes}
\usepackage[super,sort&compress,comma]{natbib} 
\usepackage[version=3]{mhchem}
\usepackage[left=1.5cm, right=1.5cm, top=1.785cm, bottom=2.0cm]{geometry}
\usepackage{balance}
\usepackage{mathptmx}
\usepackage{sectsty}
\usepackage{graphicx} 
\usepackage{lastpage}
\usepackage[format=plain,justification=justified,singlelinecheck=false,font={stretch=1.125,small,sf},labelfont=bf,labelsep=space]{caption}
\usepackage{float}
\usepackage{fancyhdr}
\usepackage{fnpos}
\usepackage[english]{babel}
\addto{\captionsenglish}{%
  
}
\usepackage{array}
\usepackage{droidsans}
\usepackage{charter}
\usepackage[T1]{fontenc}
\usepackage[usenames,dvipsnames]{xcolor}
\usepackage{setspace}
\usepackage[compact]{titlesec}
\usepackage{hyperref}
\usepackage{epstopdf}%This line makes .eps figures into .pdf - please comment out if not required.

\definecolor{cream}{RGB}{222,217,201}
\definecolor{ultramarine}{RGB}{0,32,96}
\definecolor{wrongultramarine}{rgb}{0.07, 0.04, 0.56}
\definecolor{MyRed}{rgb}{0.6350, 0.0780, 0.1840}

\begin{document}

\pagestyle{fancy}
\thispagestyle{plain}
\fancypagestyle{plain}{
%%%HEADER%%%
\renewcommand{\headrulewidth}{0pt}
}
%%%END OF HEADER%%%

%%%PAGE SETUP - Please do not change any commands within this section%%%
\makeFNbottom
\makeatletter
\renewcommand\LARGE{\@setfontsize\LARGE{15pt}{17}}
\renewcommand\Large{\@setfontsize\Large{12pt}{14}}
\renewcommand\large{\@setfontsize\large{10pt}{12}}
\renewcommand\footnotesize{\@setfontsize\footnotesize{7pt}{10}}
\makeatother

\renewcommand{\thefootnote}{\fnsymbol{footnote}}
\renewcommand\footnoterule{\vspace*{1pt}% 
\color{cream}\hrule width 3.5in height 0.4pt \color{black}\vspace*{5pt}} 
\setcounter{secnumdepth}{5}

\makeatletter 
\renewcommand\@biblabel[1]{#1}            
\renewcommand\@makefntext[1]% 
{\noindent\makebox[0pt][r]{\@thefnmark\,}#1}
\makeatother 
\renewcommand{\figurename}{\small{Fig.}~}
\sectionfont{\sffamily\Large}
\subsectionfont{\normalsize}
\subsubsectionfont{\bf}
\setstretch{1.125} %In particular, please do not alter this line.
\setlength{\skip\footins}{0.8cm}
\setlength{\footnotesep}{0.25cm}
\setlength{\jot}{10pt}
\titlespacing*{\section}{0pt}{4pt}{4pt}
\titlespacing*{\subsection}{0pt}{15pt}{1pt}
%%%END OF PAGE SETUP%%%

%%%FOOTER%%%
\fancyfoot{}
\fancyfoot[LO,RE]{\vspace{-7.1pt}\includegraphics[height=9pt]{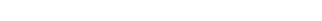}}
\fancyfoot[CO]{\vspace{-7.1pt}\hspace{13.2cm}\includegraphics{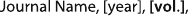}}
\fancyfoot[CE]{\vspace{-7.2pt}\hspace{-14.2cm}\includegraphics{head_foot/RF}}
\fancyfoot[RO]{\footnotesize{\sffamily{1--\pageref{LastPage} ~\textbar  \hspace{2pt}\thepage}}}
\fancyfoot[LE]{\footnotesize{\sffamily{\thepage~\textbar\hspace{3.45cm} 1--\pageref{LastPage}}}}
\fancyhead{}
\renewcommand{\headrulewidth}{0pt} 
\renewcommand{\footrulewidth}{0pt}
\setlength{\arrayrulewidth}{1pt}
\setlength{\columnsep}{6.5mm}
\setlength\bibsep{1pt}
%%%END OF FOOTER%%%

%%%FIGURE SETUP - please do not change any commands within this section%%%
\makeatletter 
\newlength{\figrulesep} 
\setlength{\figrulesep}{0.5\textfloatsep} 

\newcommand{\topfigrule}{\vspace*{-1pt}% 
\noindent{\color{cream}\rule[-\figrulesep]{\columnwidth}{1.5pt}} }

\newcommand{\botfigrule}{\vspace*{-2pt}% 
\noindent{\color{cream}\rule[\figrulesep]{\columnwidth}{1.5pt}} }

\newcommand{\dblfigrule}{\vspace*{-1pt}% 
\noindent{\color{cream}\rule[-\figrulesep]{\textwidth}{1.5pt}} }

\makeatother
%%%END OF FIGURE SETUP%%%

%%%TITLE, AUTHORS AND ABSTRACT%%%
\twocolumn[
  \begin{@twocolumnfalse}
{\includegraphics[height=30pt]{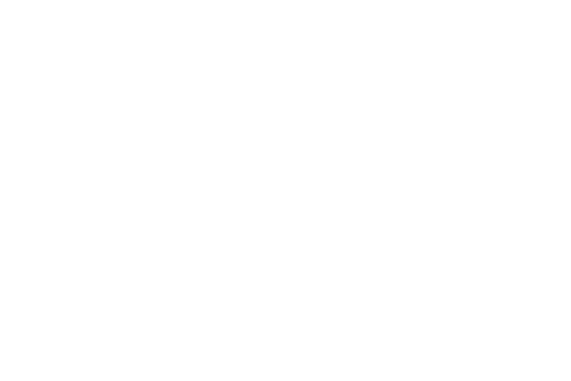}\hfill\raisebox{0pt}[0pt][0pt]{\includegraphics[height=55pt]{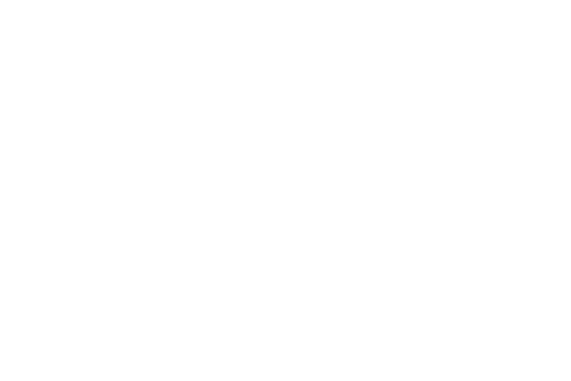}}\\[1ex]
\includegraphics[width=18.5cm]{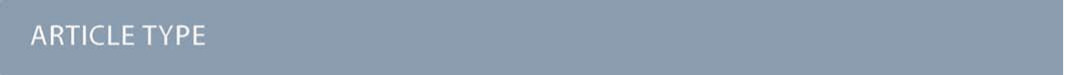}}\par
\vspace{1em}
\sffamily
\begin{tabular}{m{4.5cm} p{13.5cm} }

\includegraphics{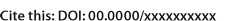} & \noindent\LARGE{\textbf{%Microscopic dynamics elucidates the origin of shear-thinning in a polydisperse dispersion of multi-lamellar vesicles$^\dag$
Microscopic structure and dynamics of shear-thinning suspensions of polydisperse, repulsive vesicles
}} \\
\vspace{0.3cm} & \vspace{0.3cm} \\

 & \noindent\large{Nikolaos Kolezakis\textit{$^{a}$}, Stefano Aime\textit{$^{b}$}, Raffaele Pastore\textit{$^{a}$},  Vincenzo Guida\textit{$^{c}$},  Gaetano D'Avino\textit{$^{a}$} and Paolo Edera$^{\ast}$\textit{$^{b}$}}\\%Author names go here instead of "Full name", etc. \ddag
\includegraphics{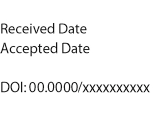} & \noindent\normalsize{We investigate the rheology, microscopic structure, and dynamics of an industrially relevant dispersion made of cationic surfactant vesicles, from dilute to concentrated conditions. 
We find that these suspensions exhibit a shear-thinning behavior at relatively low concentrations. At the microscale, this corresponds to a well-defined transition in both the structure, marked by the appearance of a peak in the static structure factor, and the dynamics, which slow down and develop a two-step decay in the correlation functions. 
This low-concentration transition is particularly surprising in light of experiments showing that for surfactant vesicles of similar composition the interactions should be purely repulsive. This leads us to propose that the observed structural and dynamic transition could arise, as an entropic effect, from the large sample polydispersity coupled to crowding. The shear-thinning behavior is thus interpreted as the nonlinear response of this transient structure to the imposed flow. Our work suggests that similar effects might be a generic feature of dense, highly polydisperse charged suspensions.}
\end{tabular}
 \end{@twocolumnfalse} \vspace{0.6cm}]
%%%END OF TITLE, AUTHORS AND ABSTRACT%%%

%%%FONT SETUP - please do not change any commands within this section
\renewcommand*\rmdefault{bch}\normalfont\upshape
\rmfamily
\section*{}
\vspace{-1cm}

%%%FOOTNOTES%%%
\footnotetext{\textit{$^{a}$ Department of Chemical, Materials and Production Engineering, University of Naples Federico II, P.le Tecchio 80, 80125, Naples, Italy.}}
\footnotetext{\textit{$^{b}$ Molecular, Macromolecular Chemistry, and Materials, CNRS, UMR 7167, ESPCI Paris, PSL Research University, 10 Rue Vauquelin, 75005, Paris, France.}}
\footnotetext{\textit{$^{c}$ Procter \& Gamble Brussels Innovation Center, 1853 Strombeek Bever, Temselaan 100, Belgium. }}
\footnotetext{\textit{$^*$ corresponding author: paolo.edera@espci.fr}}
%\footnotetext{Please use \dag to cite the ESI in the main text of the article.}
%\footnotetext{\dag~Supplementary Information available: [details of any supplementary information available should be included here]. See DOI: 10.1039/cXsm00000x/}
%additional addresses can be cited as above using the lower-case letters, c, d, e... If all authors are from the same address, no letter is required
%\footnotetext{\ddag~Additional footnotes to the title and authors can be included \textit{e.g.}\ `Present address:' or `These authors contributed equally to this work' as above using the symbols: \ddag, \textsection, and \P. Please place the appropriate symbol next to the author's name and include a \texttt{\textbackslash footnotetext} entry in the the correct place in the list.}
%%%END OF FOOTNOTES%%%

%%%MAIN TEXT%%%%
\section{Introduction}

Soft glassy materials, like dense colloidal dispersions, gels, and fiber networks, are massively exploited in the production of consumer goods. Understanding their complex rheology is crucial to controlling important features of the final products, such as their stability and shelf life, which directly impact the consumer experience. 
The rheology of these complex fluids is highly sensitive to system control parameters, such as concentration, temperature, and particle interaction, and strongly depends on the applied shear history\cite{mewis2012colloidal}, giving rise to a variety of behaviors that can be tailored to the user needs.
Among soft glassy materials relevant to the production of consumer goods, Brownian dispersions of surfactant vesicles are particularly common, with applications ranging from healthcare to laundry and cosmetics~\cite{seifert1997configurations,guida2010thermodynamics}. In the quest of minimizing the plastic use in packaging and reducing the transportation costs for consumer goods, vesicle dispersions are typically formulated at high concentration, and further diluted at need. 
Even though the rheology of concentrated colloidal suspensions has been addressed by numerous studies, both experimentally and numerically, the vast majority of these studies focus on dispersions of solid particles, with low or moderate polydispersity, while the peculiar behavior of dense, polydisperse vesicle suspensions has not received much attention~\cite{muddle1983light, seth2014origins, seth2014rheological, porpora2020understanding}.\\
\indent Here, we present a comprehensive study of a vesicle suspension of industrial interest.
Suspended in water, these vesicles carry positive surface charge. This enables to control their interaction potential through the ionic strength of the solvent, $I$\cite{leivers2018measurement}.
At low $I$, long-range Electrostatic Double Layer (EDL) interactions cause these suspensions to kinetically arrest, forming Wigner glasses, at volume fractions as small as $\phi=0.2$, well below the hard-sphere glass transition, as shown by Brownian Dynamics simulations\cite{porpora2020understanding}.
At larger $I$, the electrostatic potential at the interface is screened more efficiently by counterions, resulting in a shorter-range EDL interaction~\cite{israelachvili2011intermolecular}, whose strength and range decrease with increasing $I$~\cite{leivers2018measurement,chiruvolu1995measurement}.
This enables the preparation of liquid suspensions at larger volume fractions, with vesicle surfaces nearly in contact~\cite{porpora2020understanding}. Earlier results suggest that such dense samples display a Newtonian viscous response with zero-shear viscosity $\eta_0$, followed by shear-thinning at larger shear rates\cite{mewis2012colloidal,hunter2012physics}.
However, the Newtonian regime is often inaccessible to standard rheometry~\cite{mastersthesis,oikonomou2018design}, making it challenging to measure linear rheological properties, such as $\eta_0$, which control a number of significant spontaneous processes, including gravitational collapse and aggregation\cite{mewis2012colloidal,hunter2012physics,philippe2018glass}.
This calls for alternative experimental approaches capable of characterizing the behavior of these systems under small perturbations.\\
\indent One of the most widely exploited techniques is microrheology, which extracts linear viscoelastic moduli by probing the thermal dynamics of colloidal tracers embedded in the sample\cite{masonOpticalMeasurementsFrequencyDependent1995,masonParticleTrackingMicrorheology1997,cicutaMicrorheologyReviewMethod2007,baylesProbeMicrorheologyParticle2017,edera2017differential}. This approach assumes that the tracers' motion accurately reflects the behavior of the sample's local environment, an assumption which turns out to be especially delicate in samples with heterogeneous microstructures, including many systems of practical interest.
Alternatively, various optical techniques probe the thermal motion of the sample elementary constituents\cite{krall_internal_1998,weitz1993diffusing,pastore2017differential,cerbino2008differential,giavazzi2009scattering,edera2017differential}. Unlike microrheology, these techniques are typically unable to measure the complete viscoelastic spectrum. Yet, for liquid-like samples, they provide access to the microscopic relaxation time, $\tau$, which typically increases with packing fraction proportionally to $\eta_0$\cite{hunter2012physics,cavagna2009supercooled,janssen2018mode}. 
In this context, dense vesicle suspensions are particularly challenging to probe due to their turbidity. 
For turbid samples, Diffusing Wave Spectroscopy (DWS) is often a valid alternative, enabling the measurement of nano-scale displacements on time scales as fast as 1~$\mu$s~\cite{weitz1993diffusing}. 
For this reason, DWS has often been the technique of choice since the dawn of microrheology~\cite{masonOpticalMeasurementsFrequencyDependent1995}. 
However, for samples with complex microstructure, more sophisticated techniques are required to properly characterize nontrivial microscopic dynamics. To this end, single-scattering techniques, such as Dynamic Light Scattering (DLS) are appealing as they probe motion on a well-defined and tunable length scale, set by the scattering vector~\cite{krall_internal_1998}. Yet, DLS requires low-turbidity samples, excluding many samples of interest~\cite{walde2001enzymes,murphy2015fabric,porpora2020understanding}. For moderately turbid samples, optical microscopy techniques, such as Differential Variance Analysis (DVA) or Differential Dynamic Microscopy (DDM) are beneficial, as the reduced light coherence limits the depth of the scattering volume~\cite{aimeProbingShearinducedRearrangements2019a}, thereby mitigating the impact of multiple scattering~\cite{cerbinoDifferentialDynamicMicroscopy2022,nixon-lukeProbingDynamicsTurbid2022}. DDM, preserving the wave vector resolution, has proved to be an excellent alternative to DLS. \\
\indent In this paper, we study dense, polydisperse vesicle dispersions of industrial interest. Using rheometry, we show that these suspensions exhibit shear-thinning on the whole range of applied shear rates. Interestingly, shear-thinning persists at unexpectedly low volume fractions, much below glass transition. Combining optical microscopy, static light scattering and small angle X-ray scattering (SAXS), we show that the onset of shear-thinning coincides with a change in the sample microstructure, which was also unexpected for these repulsive suspensions at moderate volume fractions. To unveil the origin of this behavior, we use DDM to study the microscopic dynamics, and show that this microstructural transition is accompanied by a qualitative change in the dynamics, which slow down and exhibit a two-step decay in the correlation functions, occurring at volume fractions for which no caging effect was expected.\\
\indent To account for this phenomenology, reminiscent of weakly attractive colloidal suspensions, in our repulsive system, we propose the large sample polydispersity as a key ingredient. We suggest that the formation of a transient microstructure, associated with glassy-like dynamics and shear-thinning, may arise for entropic reasons, similar to what recently reported for extremely bidisperse suspensions~\cite{xuJammingDepletionExtremely2023}. As such, we suggest that a similar behavior may be a generic feature of strongly polydisperse, repulsive suspensions at moderate volume fractions.

\section{Materials and Methods}
\begin{table*}
\small
\caption{\ Samples concentration, volume fraction, and applicable experimental techniques}
\label{tbl:samples}
\begin{tabular*}{\textwidth}{@{\extracolsep{\fill}}ccccccc} % lllllll
    \hline
    concentration, $c$ [$\%$] & volume fraction, $\phi$ [$\%$]& SAXS & SALS & SLS/DLS & DDM  & Rheology \\
    \hline
    % conc  vf  & x-rays& SLS   & DLS   & DDM  & Rehology        
    0.01& 0.044 &       &   X   & X     &       &   \\
    0.1 & 0.44  & X     & X     & X     & X     &   \\
    0.3 & 1.3   & X     & X     &       &       &   \\
    1   & 4.5   & X     &       &       & X     & X \\
    2   & 9   &       &       &       &       & X \\
    4   & 18    & X     &       &       & X     & X \\
    6   & 27    & X     &       &       & X      & X \\
    8   & 36    &       &       &       &      & X \\
    10  & 45    &       &       &       &       & X \\
    11  & 49    & X     &       &       & X     & X  \\
    12  & 54    &       &       &       &       & X \\
%    13  & 53    &       &       &       &       & X \\
    \hline
  \end{tabular*}
\end{table*}
\subsection{Sample preparation}
A $13\%$ w/w aqueous dispersion of vesicles made from a double-tailed cationic surfactant (diethylester dimethyl ammonium chloride, diC18:0 DEEDMAC, mol wt = 697.5), in presence of 450 ppm of calcium chloride, was prepared using an extrusion process with an apparatus similar to that described by Corominas et al. \cite{Corominas}. The volume fraction of this starting vesicle dispersion was characterized according to the method introduced by Seth et al. \cite{seth2010dilution} and found to correspond to a volume fraction $\phi$ of $58\%$ (see Fig.~\ref{fig:vf_conc} in the SI\dag).
This dispersion was then used as a mother batch from which samples at lower concentrations were prepared through isotonic dilution. The concentrations and corresponding volume fractions of the samples are reported in Tab.~\ref{tbl:samples}, along with the techniques used for their characterization. The choice of technique for each sample was guided by the specific limitations of each method, as discussed in the following.

\subsection{Rheology}
Rheology experiments were conducted using an Anton Paar 302 MCR rheometer in rate-controlled mode with the temperature set to $25^\circ$C. To maximize the torque from the modest viscosity of the samples, a double concentric cylinder geometry was selected to measure flow curves. The shear stress, $\sigma$, was measured for shear rates, $\dot\gamma$, decreasing from $100$ to $5~s^{-1}$. The lower shear rate boundary was chosen to ensure that the torque, $T$, exceeds the sensibility limit of the instrument, $T_{min}\approx 1\ \mu N \cdot m$. 

\subsection{Small Angle X-ray Scattering}
Small Angle X-ray Scattering (SAXS) experiments have been carried out at the SWING beamline of Soleil synchrotron radiation facility.
The samples were loaded in cylindrical glass capillaries with a cross-section diameter of $1.5mm$, sealed with a bi-component epoxy glue to avoid evaporation. No sign of sedimentation or creaming was observed across several days.
Data were acquired at two different sample-detector distances to cover the widest possible wave vector range, between $10$ and $2\cdot10^4~ \mu m^{-1}$.
The background signal, $I_w$, was measured on a capillary filled with pure water and subtracted from each sample's signal, $I^R(c)$. 
The corrected signal is given by $I(c)=I^R(c)-a\cdot I_w$, where $a$ is a sample-dependent parameter adjusted between $0.95$ and $1.2$, accounting for the slight differences in each capillary diameter. The value of $a$ is estimated by collapsing the signals of all samples in the high-$Q$ limit, where the intensity scattered by the sample is negligible compared to the static background, as shown in the SI\dag~(Fig.~\ref{fig:XraysBack}).

\subsection{Optical Microscopy and DDM}\label{sec:ddm}
The samples are placed in a 125$\mu m$-thick cell made of two windows separated by a parafilm frame and sealed with epoxy glue. They are measured at room temperature with a direct optical microscope (Zeiss AXIO Scope.A1) equipped with a 10x objective and a Sony XCD-U100CR camera. 
The camera has a pixel size of $4.4\text{ x }4.4\mu m$, corresponding to $l_p=0.44\mu m$ on the object plane. Images used for DDM analysis have a resolution of $1024$x$1024$ pixels, corresponding to a lateral size of the sample area being imaged, $L=1024l_p=450.56\mu m$.
For all vesicle dispersions, two distinct image series of 120 images each are acquired at different acquisition rates, chosen to cover the range of timescales of interest. 
Acquired images are then post-processed to prevent spectral leakage effects\cite{giavazzi2017image} and analyzed using a DDM analysis code available online from {Hegelson} and coworkers\cite{ddmweb}. 
The primary output of this analysis is a set of image correlation functions $D(Q,\Delta T)$, for scattering vectors $Q$ ranging from $2\pi/L$ to $\pi/l_p$.
When the contribution of the static background to the recorded intensity can be neglected, the image correlation function can be normalized using the image static structure function $D_0(Q)=\langle|\tilde{I}_p(Q,T)|^2\rangle_T$, where $\tilde{I}_p$ is the Fourier transform of the intensity detected by the $p$-th pixel of the camera, and $\langle\cdots\rangle_T$ denotes averaging over time. If the camera noise can be neglected, the intermediate scattering function (ISF) is then obtained by taking $f(Q, \Delta T)=1-D(Q,\Delta T)/D_0(Q)$, as detailed in the SI\dag. 
Finally, ISFs extracted from videos recorded at different rates are merged into a unique set of correlation functions.\\
\indent We also exploit DDM analysis to obtain information on the static intensity scattered by the sample, $I(Q)$. To this end, we model $D_0(Q)$ as the product of $I(Q)$ times the microscope transfer function, $T(Q)$~\cite{giavazzi2009scattering}, neglecting the contribution of the static background intensity, which is negligible in the experimental conditions considered in this paper. Because $T(Q)$ depends uniquely on the imaging conditions, which are kept constant for all the measured samples, we correct for it by normalizing the image static structure functions by that measured for the most diluted sample: $I(c,Q)/I(c_0,Q)=D_0(c,Q)/D_0(c_0,Q)$, where $c_0=0.1\%$.
For a monodisperse colloidal suspension, $I(Q)$ would be expressed as a product of the particle form factor, $P(Q)$, which would be independent of concentration, times the structure factor, $S(Q)$, which would tend to $1$ with decreasing concentration. In that case, we expect that $S(Q)\approx I(c,Q)/I(c_0,Q)$. Our vesicle dispersion is highly polydisperse, in which case we use the same data analysis protocol to obtain the so-called measurable structure factor, $S^M(Q)\approx I(c,Q)/I(c_0,Q)$~\cite{zemb_neutron_2002}

\section{Results}
\subsection{Rheology}
\begin{figure*}
 \centering
 \includegraphics[width=1.0 \linewidth ]{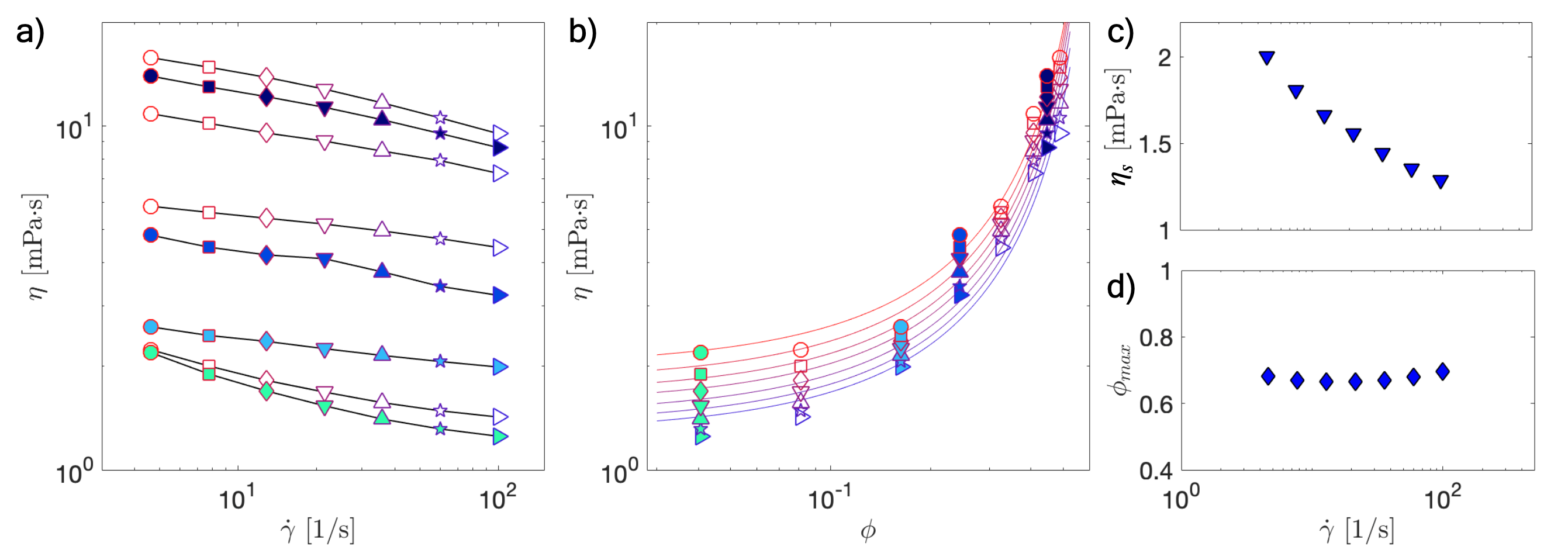}
 \caption{
 \textbf{Rheological characterization.} 
 Symbols: Steady-state viscosity as function of the shear rate (a) and of the volume fraction. (b) 
 Full symbols: samples at $\phi=4.5\%$ (green), $18\%$ (cyan), $27\%$ (blue), $49\%$ (black), investigated in more detail in the rest of the paper. 
 Solid lines in (b): fit through the Krieger-Dougherty (Eq.~\ref{eq.krieger}),
 (c) zero concentration viscosity $\eta_s$, and (d) maximum volume fraction $\phi_{max}$.
 }
\label{fig:vesicles}
\end{figure*}

We measure the rate-dependent viscosity, $\eta(\dot\gamma)$, of samples at different concentrations, between $\phi=4.5\%$ and $54\%$. For all samples, we find that $\eta$ increases as $\dot\gamma$ is decreased, indicating shear-thinning, as illustrated in Fig.~\ref{fig:vesicles}a.
At constant rate, we find that $\eta$ increases with $\phi$, in a way that is well described by the Krieger-Dougherty equation \cite{krieger1959mechanism}, 
\begin{equation}
    \eta=\eta_s\cdot (1- \phi/\phi_{max})^{-2.5\cdot \phi_{max}}
    \label{eq.krieger}
\end{equation}
as shown in Fig.~\ref{fig:vesicles}b.\\
\indent In its original formulation, this model describes the volume fraction dependence of the zero-shear viscosity, as a function of the solvent viscosity, $\eta_s$, and a phenomenological threshold volume fraction, $\phi_{max}$, at which the viscosity is predicted to diverge. 
However, here we find that while $\phi_{max}$ has a consistent value of about 70\%, in line with the assumption of short-range, hard-sphere-like interparticle repulsion, $\eta_s$, as determined by fitting our data with Eq.~\ref{eq.krieger}, decreases with increasing $\dot\gamma$ (see Fig.~\ref{fig:vesicles}c). 
This suggests that shear-thinning does not arise from simple crowding, but rather from a structural relaxation time $\tau_R \gg \dot\gamma^{-1}$, whose origin is unclear \cite{cheng2011imaging,xu2013relation}. 
To unveil the origin of this shear-thinning behavior, we study the microscopic structure and dynamics of these vesicle suspensions.

\subsection{Microscopic structure \label{sec:struct}}
\begin{figure}[ht!]
\centering
  \includegraphics[width=0.9 \linewidth]{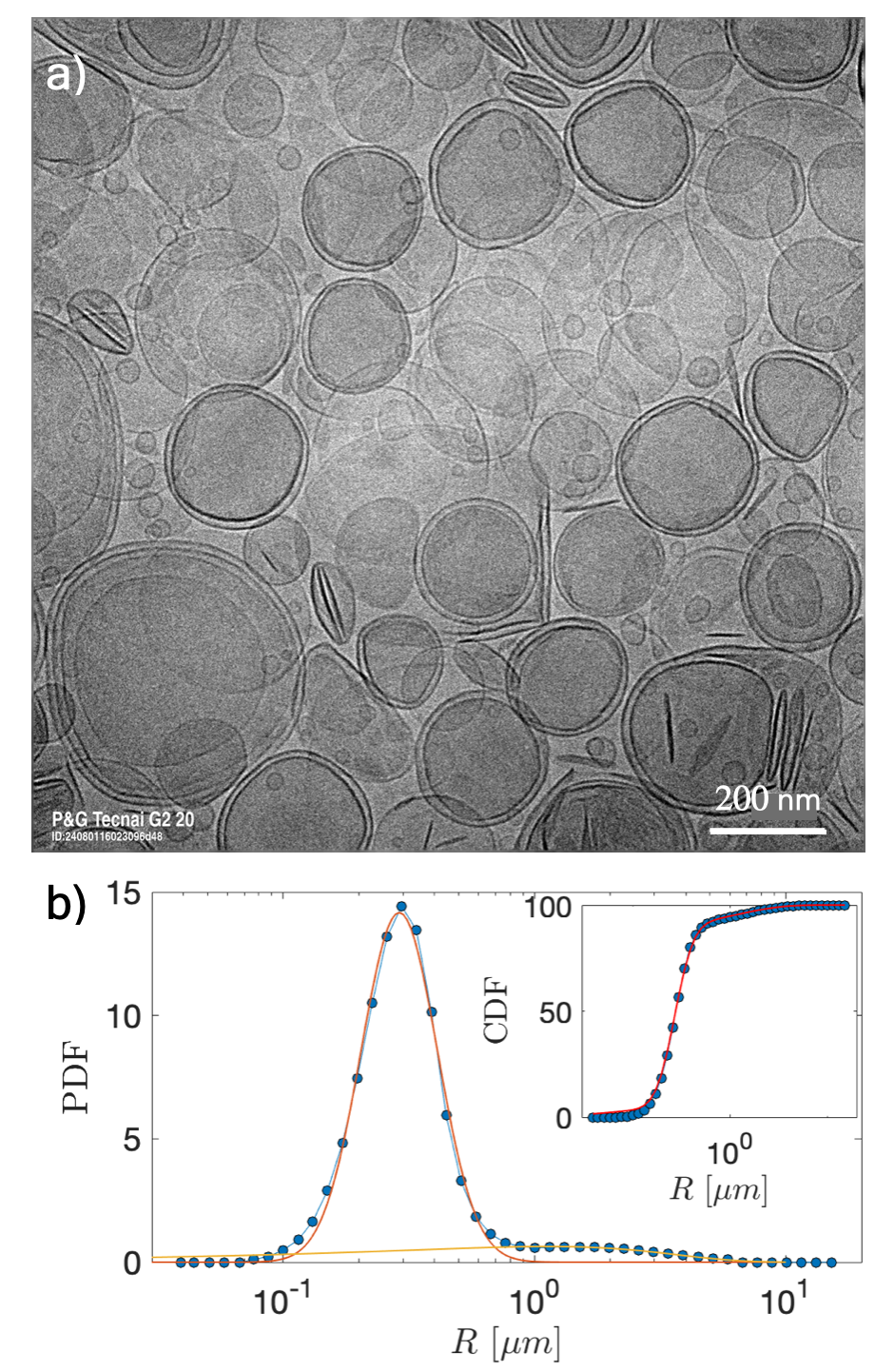}
 \caption{
 a) Cryo-TEM image of the dispersion with $\phi=58\%$ showing the multilayer structure of vesicles, the extreme polydispersity of shape and size, and the fluctuations from spherical form that are due to the membrane softness. b) Size distribution sampled by LA-950 particle size analyzer (see main text). Inset: Cumulative particle size distribution.}
\label{fig:XRAYS}
\end{figure}

\paragraph*{Cryo-TEM and particle sizing}
As an initial characterization of the sample's microstructure, we examine a concentrated vesicle suspension ($\phi\simeq58\%$) through cryo-TEM imaging \cite{saha2017impact}. 
We find that the system is highly polydisperse, as shown in Fig.~\ref{fig:XRAYS}a. \\% Additionally, we observe that several vesicles exhibit significantly non-spherical shapes, suggesting that their bending rigidity is of the order of $k_B T$.\\
\indent To better characterize sample polydispersity, we sample it using a commercial particle size analyzer based on static light scattering from individual particles flowing through the scattering volume (model LA-950, from Horiba). We obtain the result shown in Fig.~\ref{fig:XRAYS}b, indicating that vesicles with size between $100nm$ and $1\mu m$ contribute to about $90\%$ of the volume fraction, with a median diameter of about $280nm$, in agreement with the qualitative estimate based on cryo-TEM, and a polydispersity of about $160\%$, defined as the variance of the size distribution, normalized by its squared mean.\\ 
\indent Particle softness and polydispersity typically shift the glass transition towards larger volume fractions\cite{farrClosePackingDensity2009}, further ruling out the hypothesis that the observed shear-thinning behavior results from the onset of caging.

\begin{figure}[t!]
\centering
  \includegraphics[width=1.0 \linewidth]{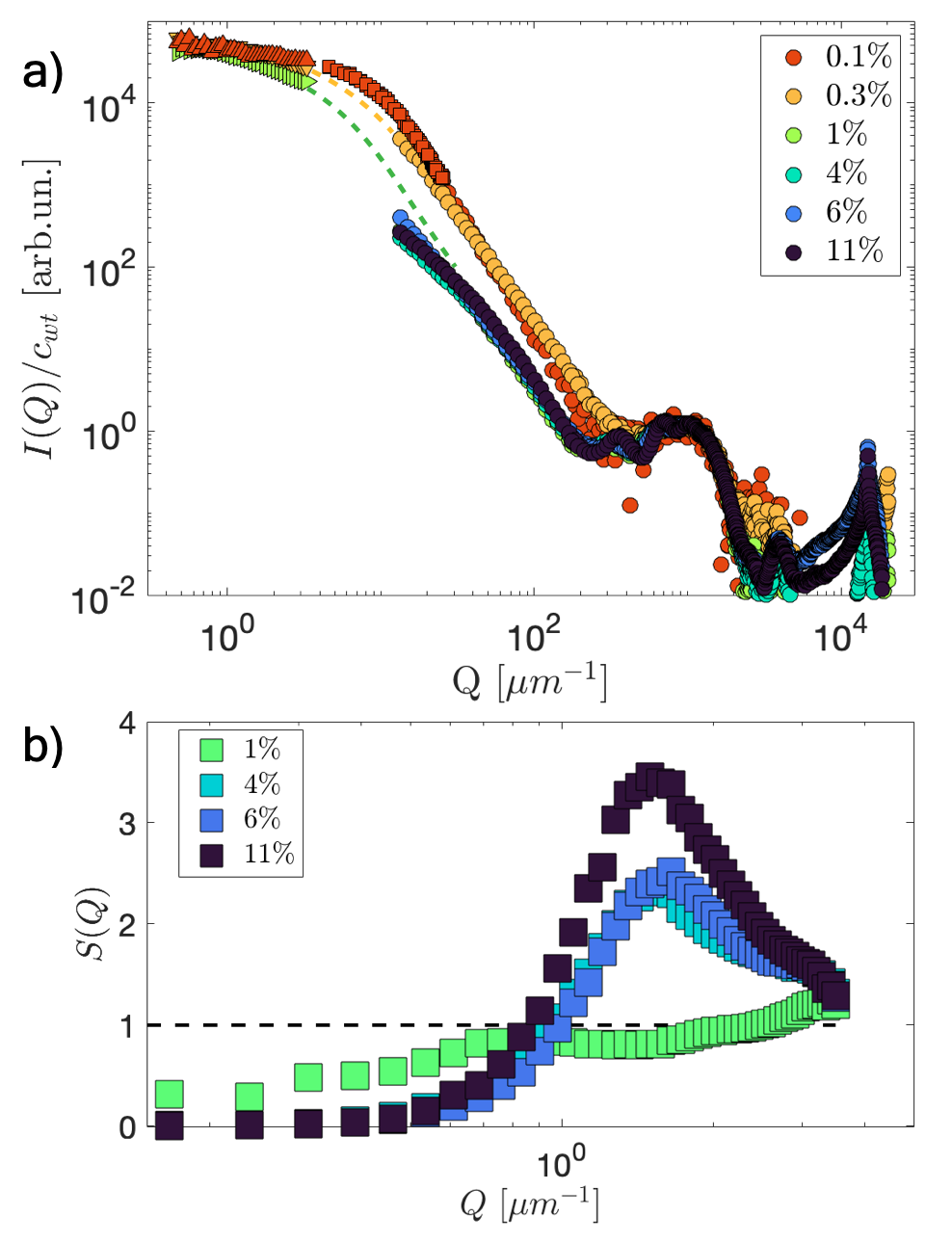}
  \caption{Microscopy, static indicators. 
  a) Circles: SAXS intensity for different mass concentrations, $c$, as specified in the legend. Triangles: SALS intensity for the most diluted samples, $c\leq 1\%$. Squares: SLS experiments at large angles for $c=0.1\%$. Dashed lines are a guide to the eye, connecting SAXS and SALS data in the range of intermediate $Q$, where SLS is not applicable. 
  b) Static structure functions, obtained from normalizing the high concentrations image structure functions with the low-concentration image structure function: $S(Q;c)= D_0(Q;c)/D_0(Q;c=0.1\%)$. 
  }
  \label{fig:MultiScatt1}
  \end{figure}

\paragraph*{Static Light and Small Angle X-ray Scattering}
To more properly characterize the $\phi$-dependent structure of such a polydisperse sample, we complement Cryo-TEM measurements with Small Angle X-ray scattering (SAXS).
%\indent Because of the weak scattering cross-section of X rays, SAXS data could cover a wide range of sample concentrations, from the dilute regime ($c=$0.01\%) up to the largest volume fraction studied in this work ($c=11\%$, corresponding to $\phi\simeq $49\%). 
Comparing samples at different concentrations, we find that the scattered intensities, $I(Q)$, almost collapse on a $\phi$-independent mastercurve when normalized by the concentration of surfactant, $c$, as shown in Fig.~\ref{fig:MultiScatt1}a. More precisely, we obtain a nice collapse for large scattering vectors, %where $I(Q)/c$ exhibits a broad peak around $Q\simeq 1 nm^{-1}$, which we interpret as the signal of the surfactant double layer of thickness around 5~nm \cite{schmiedelDeterminationStructuralParameters2001}, and a second sharp peak at $Q= 12 nm^{-1}$ \SA{check}, which we interpret as the signature of the lateral ordering of surfactant molecules in the bilayer \SA{REF}.
while for smaller $Q$, we identify two distinct families of curves: one for lower concentrations, with $c< 1\%$, characterized by an extended power-law regime, $I\propto Q^{-3}$ for $0.01<Q<0.1 nm^{-1}$, presumably arising from the large sample polydispersity, and another for higher concentrations, qualitatively similar to the first one, but shifted towards smaller $Q$. For the lowest concentrations, samples are optically clear, which enables the extension of SAXS measurements at smaller scattering vectors by Small Angle Light Scattering (SALS) and standard Static Light Scattering (SLS). The resulting mastercurve exhibits a plateau at small $Q$ and a broad crossover around $Q_c\approx 8\mu m^{-1}$, indicating a characteristic vesicle size $R= 2\pi/Q_c \simeq 0.8 \mu m$, as shown in Fig.~\ref{fig:MultiScatt1}a. 
This value is compatible with the size distribution sampled above, considering that the scattered intensity is dominated by the largest vesicles.\\
\indent For larger surfactant concentrations, $c \geq 1\%$, we find that $I(Q)$ departs from the low-concentration behavior, shifting towards lower scattering vectors. 
One possible interpretation of this effect is a change in microstructure due to the formation of permanent or transient vesicle aggregates. 
To test this hypothesis, we would need to examine the sample's structure at small scattering vectors to identify the onset of a measurable structure factor. Unfortunately, static light scattering cannot probe these low $Q$-values for concentrations higher than $1\%$ due to sample turbidity. To overcome this limitation, we extract the static structure factor from bright-field microscopy videos, as described in section \ref{sec:ddm}.

\begin{figure}[h!]
    \centering
    \includegraphics[width=0.95\columnwidth]{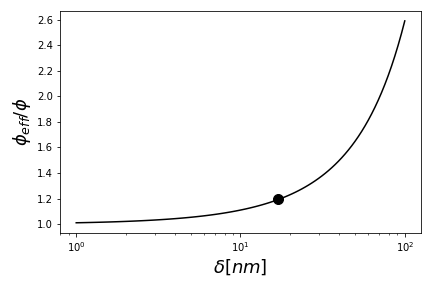}
    \caption{Solid line: relative increase in the effective volume fraction, $\phi_{eff}$, obtained if each particle in the size distribution had its radius increased by an amount $\delta$. The black circle highlights the interaction distance, $d^*=17nm$, computed in the text, which corresponds to $\phi_{eff}\approx 1.2\phi$.}
    \label{fig:effective_volfr}
\end{figure}

\paragraph*{Microscopy}
We find that for concentrations above 1\%, the measurable structure factor $S^M(Q)$ exhibits a peak around $Q\simeq1.5\ \mu m^{-1}$, corresponding to a length scale $2\pi/Q \simeq 4 \mu m$, a few times larger than the average particle size extracted by $I(Q)$ in the dilute limit. In addition, we find that $S^M(Q)$ decreases with $\phi$ in the $Q\rightarrow 0$ limit. 
These features could be indicative of an incipient glass transition~\cite{percusAnalysisClassicalStatistical1958}. However, we find that the peak appears at a low volume fraction, below $20\%$, and exhibits a weak dependence on $\phi$, only increasing by $50\%$ as $\phi$ is nearly tripled, as shown in Fig.~\ref{fig:MultiScatt1}b. These features contrast with what is typically found at the colloidal glass transition. 
To further rule out the hypothesis that a soft, long-range repulsive potential may be the cause of an early and more gradual glass transition, we consider once more the size distribution of Fig.~\ref{fig:XRAYS}b, and compute the relative increase in the effective volume fraction that would be obtained if the long range repulsion was forcing the vesicle membranes to stay a distance $d^*$ apart. The result, shown in Fig.~\ref{fig:effective_volfr} (SI\dag), exhibits only a mild increase for $d^*$ in the nanometer range. A realistic calculation, detailed in the Appendix, yields $d^*\approx 15nm$, for which we obtain an effective volume fraction, $\phi_{eff}\simeq 1.2\phi$, well below the increase needed to justify an approaching glass transition at~ $\phi< 20\%$.\\
\indent An alternative hypothesis accounting for this low-$\phi$ structuring is that vesicles aggregate due to weak attractive interactions. However, direct measurements of the interaction potential between two surfactant bilayers, similar to those composing our system, indicate that the interaction between individual vesicles is purely repulsive \cite{leivers2018measurement}. 
Another mechanism causing structure formation in repulsive colloids is depletion: effective attraction between large, repulsive colloids dispersed in a solution of small, non-adsorbing polymer chains. When the degrees of freedom of polymers are coarse-grained in an effective solvent medium, the suspended colloidal particles turn out to be subjected to an effective attractive interaction, of entropic origin~\cite{maoDepletionForceColloidal1995}. 
The same concept has been exploited in the case of bidisperse colloidal suspensions, when the size of the two colloidal populations is significantly different~\cite{bibetteDepletionInteractionsFluidsolid1990,pednekarBidispersePolydisperseSuspension2018}.\\
\indent Here, we propose that a similar entropic effect is at play in our vesicle suspension, due to its large polydispersity. Our suspension is not bidisperse, therefore no clear distinction can be made between small particles being coarse-grained in the effective medium and large particles feeling an effective attraction. 
Yet, we propose that in our strongly polydisperse, repulsive suspension, entropy may be maximized by locally clustering particles of similar size, a mechanism analogous to size segregation in granular materials~\cite{mobiusSizeSeparationGranular2001}. 
To test this hypothesis, we study the microscopic dynamics of the samples at different concentrations.

\subsection{Microscopic dynamics}
To mitigate the effect of multiple scattering, we decide to probe the microscopic dynamics of our vesicle suspensions through DDM, following the procedure detailed in Sec.~\ref{sec:ddm}.

\begin{figure}[t]
\centering
  \includegraphics[width=0.95 \linewidth]{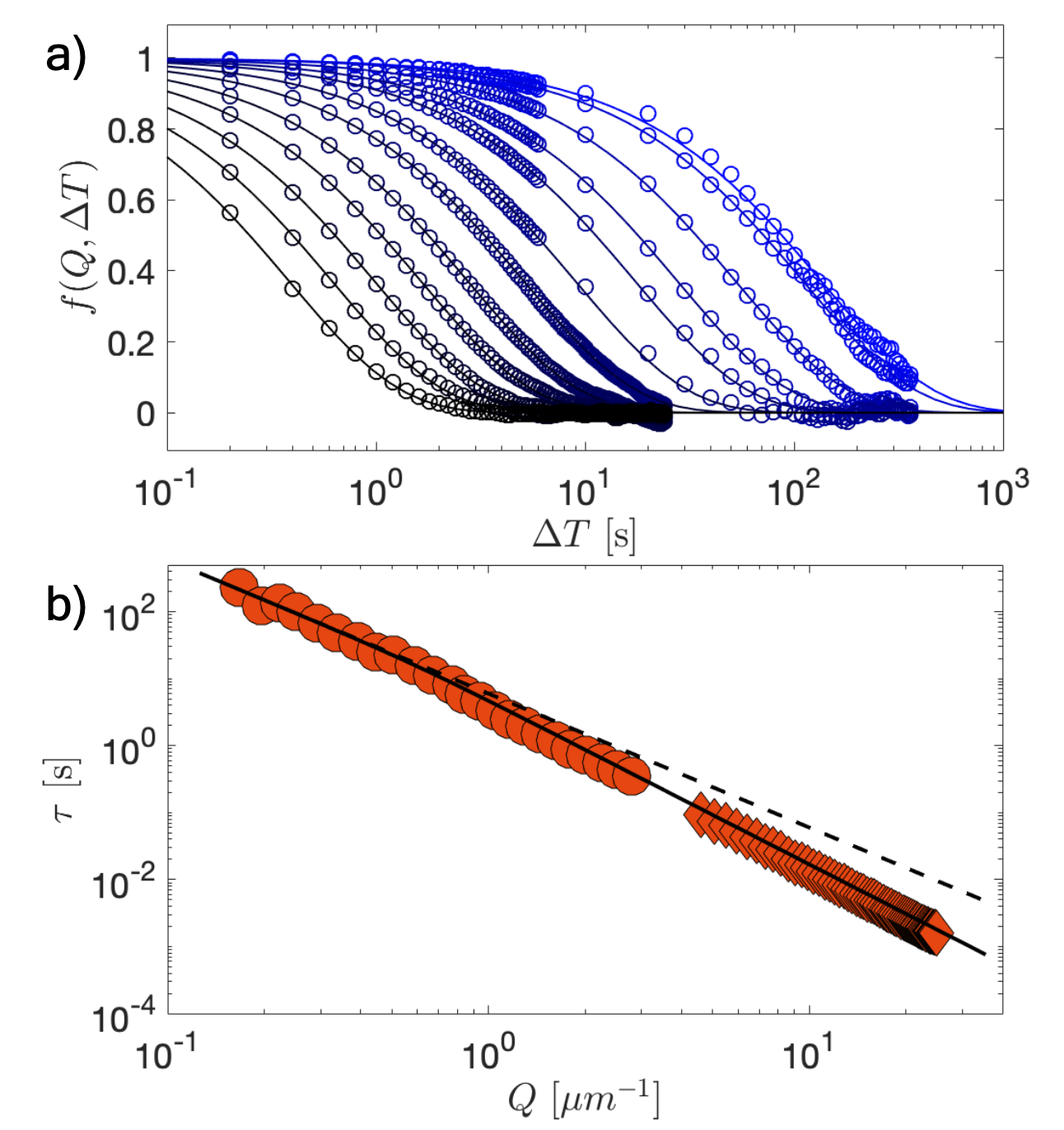}
  \caption{
  \textbf{Microscopic dynamics in the dilute samples}.
  (a) Symbols: ISFs in the dilute regime ($c=0.1\%$) for $Q$ increasing from $0.17$ (blue) to  $2.79\ \mu m^{-1}$ (black). Lines: stretched exponential fits with shared stretching exponent $\beta=0.825$. (b) Average relaxation times from DDM (circles) and from DLS (diamonds).
  Dashed line: diffusive fit $\tau=1/(D Q^2)$ to data at $Q<0.6\ \mu m^{-1}$, yielding $D\simeq 0.17 \frac{\mu m^2}{s}$. Solid line: sub-diffusive fit $\tau=A Q^{-p}$ to data at $Q>0.7\ \mu m^{-1} $, yielding $A= 4.25 $ and $p=2.45$.
  %\SA{Use larger axis labels and titles, larger symbols in panel a, larger aspect ratio to adapt to the column width, add units for $\tau$ and $\Delta T$}
  }
\label{fig:Dilut}
\end{figure}

\paragraph*{Dilute regime}
For samples in the dilute regime ($c<1\%$), $f(Q, \Delta T)$ displays a single decay, well-fitted by a stretched exponential, $f=\exp[-(\Delta T/\tau)^\beta]$, with a $Q$-independent stretching exponent, $\beta=0.825$, as illustrated in Fig.~\ref{fig:Dilut}a for $c=0.1\%$. From these fits, we extract the average relaxation time, $\tau(Q)$, shown as circles in Fig.~\ref{fig:Dilut}b. 
For small scattering vectors, $Q<1\mu m^{-1}$, we find that $\tau(Q)\propto Q^{-2}$, consistent with Brownian motion, where $\tau(Q) = 1/(D Q^2)$, with $D$ being the diffusion coefficient wich is expected to obey to the Stokes-Einstein relationship, $D=k_B T/(6\pi\eta_0 R)$. A fit to the data in this low-$Q$ regime yields $D\simeq 0.18 \mu m^2/s$, corresponding to $R=1.2 \mu m$, in good agreement with the vesicle size inferred from Fig.~\ref{fig:MultiScatt1}a.\\
\indent For larger scattering vectors ($Q>1\mu m^{-1}$), we find that the Brownian diffusion model slightly overestimates the value of $\tau(Q)$ obtained experimentally, as shown by the dashed line in Fig.~\ref{fig:Dilut}b. In this $Q$ range, a sub-diffusive model, 
$\tau(Q)\propto Q^{-p}$ with $p\simeq 2.4 $, seems more appropriate to fit the data. 
To confirm this trend, we extend these results by DLS experiments at large scattering angles, using the measured correlation functions to extract intermediate scattering functions fully compatible with those measured by DDM, as shown in the SI\dag. Relaxation times, extracted by stretched exponential fits, extend the probed $Q$ range by about a decade, confirming that the dynamics in the large-$Q$ regime is sub-diffusive, as illustrated by diamonds in Fig.~\ref{fig:Dilut}b. 
We find that the onset of this transition from diffusive to sub-diffusive scaling occurs at a value of $Q$ that roughly coincides with the departure of $I(Q)$ from its low-$Q$ Guinier regime. This suggests that the observed sub-diffusive dynamics concerns objects smaller that the average particle size. As such, it may be attributed either to polydispersity, or to additional relaxation modes associated with shape fluctuations, that are expected for flexible vesicle membranes in the regime $Q R > 1$.

\paragraph*{Impact of multiple scattering with increasing concentration}

As vesicle concentration increases, the growing turbidity restricts the applicability of single-scattering techniques. Wide-angle DLS is effective only for concentrations up to $c=0.1\%$, while SALS, benefiting from reduced sample thickness, can extend this range to $c=1\%$. For concentrations beyond $1\%$, DDM remains the sole technique probing single-scattered light. To evaluate the effect of multiple scattering in DDM experiments at the highest concentrations, we analyze the histogram of the rescaled image intensity, $P(\tilde{I})$, as defined in the SI\dag. Up to $c=6\%$, $P(\tilde{I})$ is well-captured by a Gaussian distribution, as expected for single scattering from a random distribution of phase objects in the scattering volume \cite{giavazzi2009scattering}. At $c=11\%$, the histogram becomes slightly skewed, suggesting that the image formation process is no longer strictly linear, as shown in Fig.~\ref{fig:MultiScatt2}a (SI\dag). 
For this concentration, we assess the $Q$-dependent impact of weak multiple scattering. We expect greater impact at small $Q$ due to the larger depth of focus \cite{aimeProbingShearinducedRearrangements2019a}, and we aim at defining a viable $Q$ range where multiple scattering can still be neglected. To this end, we analyze the rescaled image static structure function, $\tilde{D}_0=a(c)\cdot D_0$, defined in the SI\dag. For $Q>0.9\ \mu m^{-1}$, the data measured at different concentrations collapse nicely, while at lower $Q$, a weak excess in the scattered signal is observed for the most concentrated sample, as shown in Fig.~\ref{fig:MultiScatt2}b (SI\dag). Based on this observation, we limit the analysis of the $c=11\%$ sample to the high-$Q$ regime, $Q>0.9\ \mu m^{-1}$, while for the rest of the samples, we study the whole accessible $Q$ range.

\paragraph*{Concentrated regime}

\begin{figure}[h!]
    \centering
    \includegraphics[width=0.95\linewidth]{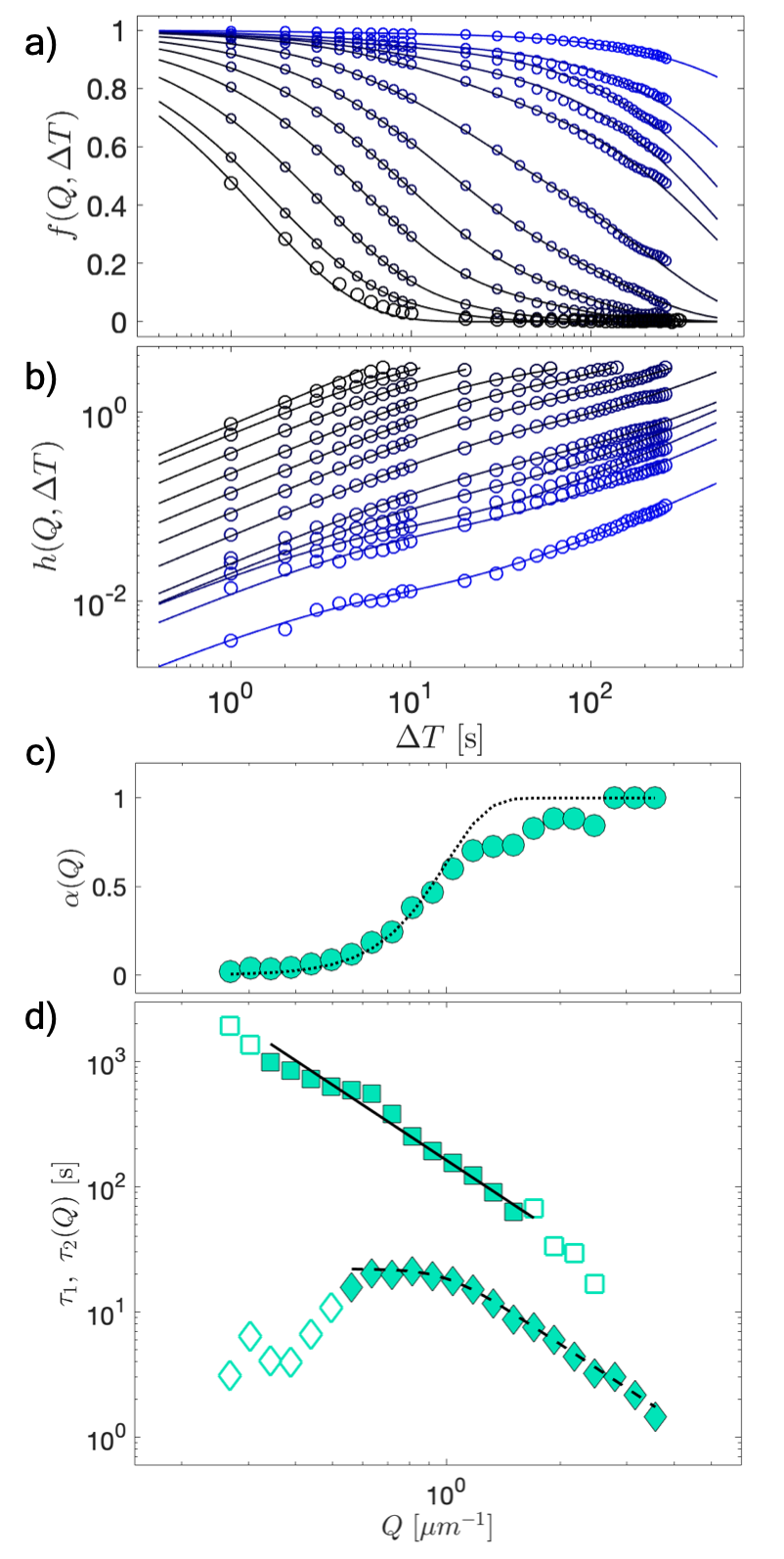}
    \caption{
    \textbf{Microscopic dynamics for concentrated samples}.
    a) Symbols: ISFs of the sample with $c=4\%$ for $Q \in [0.2, 3.5]\mu m^{-1}$, increasing from blue to black. Lines: double stretched exponential fits (Eq.~\ref{eq:isf_double}), with shared stretching exponents $\beta_1=0.8$ and $\beta_2=0.9$.
    b) $h=-\ln(f)$ to highlight the presence of two distinct processes. Data and fits are the same as in panel a, truncated at $h<3$ ($f<0.05$), beyond which $h$ is dominated by the imperfect normalization of the ISF that introduces an uncertainty in the $f=0$ baseline.
    c-d) Fitting parameters: $\alpha(Q)$ (circles), $\tau_1(Q)$ (diamonds), and $\tau_2(Q)$ (squares), extracted from data in panels (a-b). Open diamonds correspond to $\alpha<0.1$, open squares correspond to $\tau_2>10^3\ s$ or to $\alpha>0.8$. Dotted line: % \SA{perhaps make it dotted, or dash-dotted, to simplify the description} 
    fit to $\alpha(Q)=1-\exp[-(Q/Q^*)^4]$, with $Q^*=1 \ \mu m^{-1}$.
    Dashed line: fit to $\tau_1(Q)=\tau_b/[1+(Q/Q^*)^{2\cdot\epsilon}]^{1/\epsilon}$, with $\tau_b=22\ s$, $\epsilon= 4 $, and $Q^*=1\ \mu m^{-1}$. Solid line: fit to a diffusive scaling $\tau_2=1/(D_{slow} Q^2)$ with $D_{slow}=7.4\cdot 10^{-3} \mu m^2/s$.
    }
    \label{fig:Dyn_04}
\end{figure}

To highlight the qualitative features of the dynamics observed in the concentrated regime, we consider the case of $c= 4\%$, corresponding to a vesicle volume fraction of $\phi=18 \%$.
The most striking difference from the dilute case, shown in Fig.~\ref{fig:Dilut}a, is that in an intermediate range of scattering vectors, $f(Q, \Delta T)$ displays a two-step decay. While this feature is not easily detectable from Fig.~\ref{fig:Dyn_04}a, this behavior can be clearly highlighted by taking the logarithm of the ISF, $h(Q,\Delta T)=-\ln[f(Q,\Delta T)]$, as in Fig.~\ref{fig:Dyn_04}b. We find that $h$ grows as $\Delta T^{\beta_1}$ for $Q>2.4 \ \mu m^{-1}$, while at lower $Q$, it exhibits two distinct power-law regimes separated by a pseudo-plateau, which becomes more pronounced as $Q$ decreases, as shown in Fig.~\ref{fig:Dyn_04}b.\\
\indent We find that in the whole range of $Q$ accessed by DDM, $f(Q, \Delta T)$ is well-described by a sum of stretched exponential decays with fixed stretching exponents, $\beta_1=0.8$ and $\beta_2=0.9$, and $Q$-dependent relaxation times $\tau_1(Q)$ and $\tau_2(Q)$:

\begin{equation}
        f(Q, \Delta T)=\alpha(Q) e^{-\left[\frac{\Delta T}{\tau_1(Q)}\right]^{\beta_1}} + \left[1-\alpha(Q)\right] e^{-\left[\frac{\Delta T}{\tau_2(Q)}\right]^{\beta_2}} \, ,
\label{eq:isf_double}
\end{equation}
\noindent where $\alpha$ is the $Q$-dependent relative amplitude of the two decays. Fitting Eq.~\ref{eq:isf_double} to our experimental data, we observe that $\alpha(Q)$ follows a sigmoidal shape, describing a gradual transition between two relaxation modes, around a characteristic scattering vector $Q^*$, as shown in Fig.~\ref{fig:Dyn_04}c. This suggests that the thermal relaxation of these dense suspensions is characterized by two distinct dynamics, reminiscent of glassy systems \cite{pastore2022multiscale}: a fast relaxation process, entailing small-scale displacements that dominate the signal at large $Q$, and a slow relaxation process that involves larger-scale motion over longer times. This slow dynamics does not contribute significantly to the decay at large $Q$, which is dominated by the fast, small-scale motion, and becomes increasingly important at smaller $Q$, for which the length-scale, $\xi$, of fast displacements is too small to produce a full decorrelation. 
The threshold value $Q^*$ thus represents the characteristic length scale of the fast dynamics. By fitting our data to the empirical law, $\alpha(Q)=1-\exp\left[-(Q/Q^*)^4\right]$, we obtain $Q^*=1\ \mu m^{-1}$, which is close to the structural peak observed in $S^M(Q)$, shown in Fig.~\ref{fig:MultiScatt1}b. This suggests that the fast relaxation mode involves displacements as large as $\xi=2\pi/Q^*\simeq 1.5\ \mu m$, which corresponds approximately to the length scale of structural heterogeneity.\\
\indent The nature of these dynamics can be investigated by analyzing the $Q$-dependence of the two relaxation times. We find that the longest relaxation time, $\tau_2$, is diffusive, and well-described by a diffusion coefficient, $D_{slow}=1/(\tau_2 Q^2)$ with $D_{slow}=7.4\cdot 10^{-3} \mu m^2/s$. By contrast, $\tau_1(Q)$ exhibits a transition between diffusive-like dynamics for $Q>Q^*$, and a small-$Q$ regime where $\tau_1$ depends more weakly on $Q$, apparently reaching a plateau value below $Q^*$. This regime is harder to characterize, due to the small relative amplitude of the fast decay implied by the decaying $\alpha\rightarrow 0$, as shown in Fig.~\ref{fig:Dyn_04}d. For simplicity, we capture this transition with the empirical relationship, $\tau_2(Q)=\tau_b/[1+(Q/Q^*)^{2 \epsilon}]^{1/\epsilon}$, describing a $Q$-independent relaxation time, $\tau_b$, for $Q\ll Q^*$, and a diffusive relaxation, $\tau_2=1/(D_{fast} Q^2)$, for $Q\gg Q^*$, and with an additional parameter, $\epsilon \simeq 4$, controlling the sharpness of the transition between the two regimes. We fit this model to the data for $Q>0.5\ \mu m^{-1}$, corresponding to $\alpha>0.1$, thereby disregarding $\tau_1$ extracted from correlation functions where the contribution of the fast dynamics accounts for less than $10\%$ of the total decay, as such condition prevents a robust estimation of the fit parameters. We find $Q^*=1\ \mu m^{-1}$, equal to the value extracted from fitting $\alpha(Q)$, and $\tau_b=22\ s$. 
The short-time diffusivity, $D_{fast}=1/(\tau_b {Q^*}^2)=0.045 \mu m^2/s$, describes fast diffusion on the heterogeneity length scale, $1/Q^*$, in a time scale set by $\tau_b$. We find a sizable slow down, about a factor 4, relative to the Brownian diffusion measured in dilute samples.\\
\indent We interpret the coexisting fast and slow dynamics, unveiled by DDM, as yet another indication that in this range of concentrations the sample changes its microstructure, by locally segregating vesicles of similar sizes, in microscopic domains of a few microns in size. Within this framework, fast and slow dynamics could correspond to motion of particles within one domain, and the relaxation of the domains themselves, respectively. The slow down of $D_{fast}$ relative to the free diffusion coefficient suggests that the diffusion of individual vesicles is hindered by crowding, which is a necessary condition for the proposed entropic size segregation to take place. In addition, the relatively small separation of the two timescales, less than one decade, suggests that the entropic drive towards size segregation is rather weak, which is compatible with the experimental evidence that the length scale of structural heterogeneity is just a few times larger than the size of individual vesicles.

\paragraph*{Concentration dependence}

\begin{figure}[t]
\centering
  \includegraphics[width=0.95\linewidth]{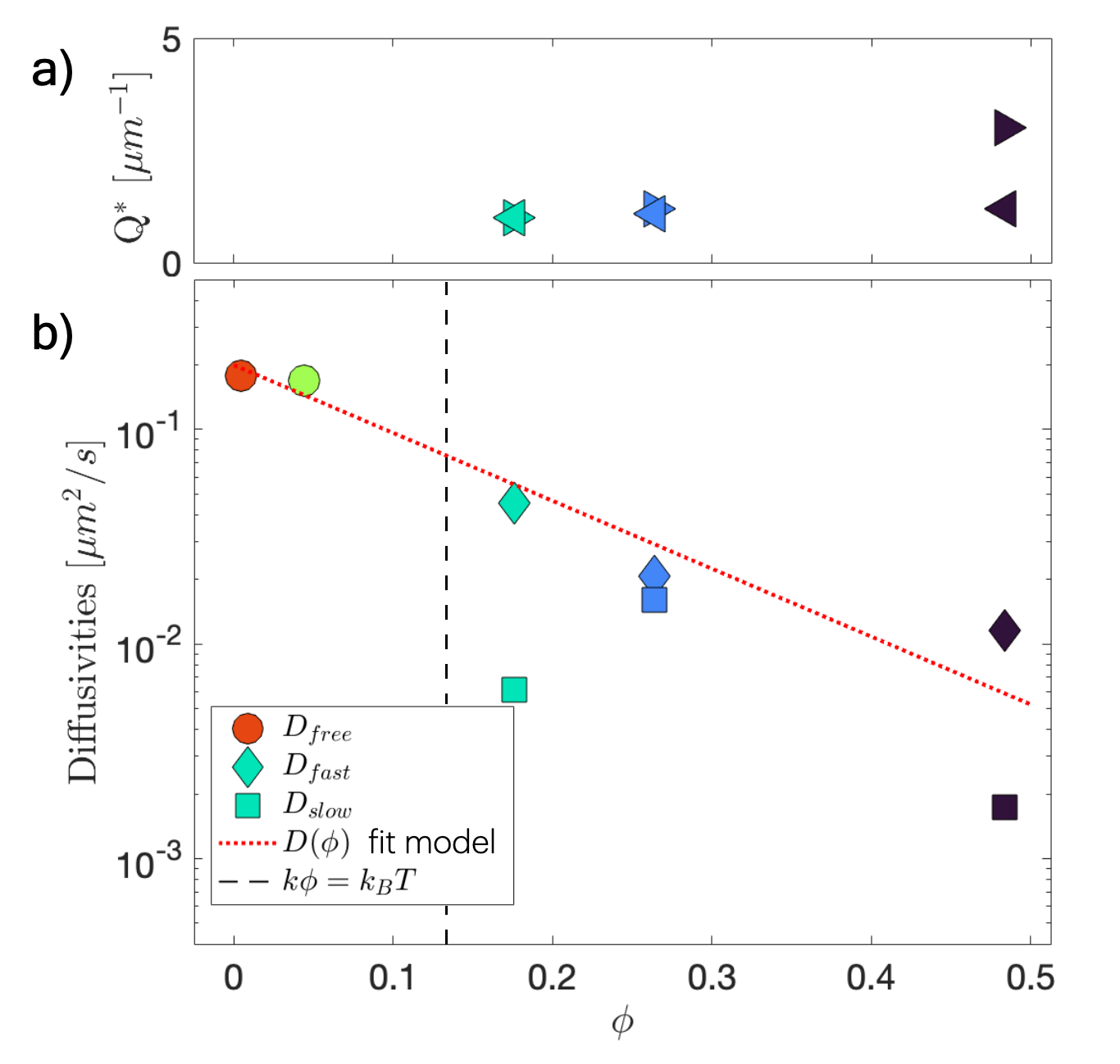} 
  \caption{
  \textbf{Concentration dependence of microscopic dynamics}. 
  a) Characteristic wave vector $Q^*$ for $c=4\%$ (cyan), $c=6\%$ (blue), and $c=11\%$ (black), as estimated from the dependence of the fast decay amplitude $\alpha(Q)$ ($\triangleright$), and from the dependence of the fast decay time $\tau_1(Q)$ ($\triangleleft$).  
  b) Circles: free diffusion coefficients in the diluted regime for $c=0.1\%$ (red) and $c=1\%$ (green). Diamonds and squares: fast and slow diffusion coefficients in the concentrated regime, respectively. 
  Red line: fit to $D_{free}$ and $D_{fast}$ using the empirical model $D(\phi)=D_0 \exp(-k\phi)$, with $k= 7.3$, as discussed in the text. Black dashed line: predicted threshold for microstructural transition.}
\label{fig:RelaxTimes}
\end{figure}

To further characterize the stability of the transient microstructure, we extend the above analysis to the dynamics of even more concentrated samples, with $c=6\%$ and 11\%. We obtain similar results to those shown above, with ISFs well-fitted by Eq.~\ref{eq:isf_double}, yielding stretching exponents reported in Tab.~\ref{tbl:stretch} (SI\dag) and $Q$-dependent fitting parameters exhibiting the same qualitative features as the sample at $c=4\%$, as shown in Fig.~\ref{fig:Concentr} (SI\dag).
Comparing the fit results at different concentrations, we find that $Q^*$ is either constant or weakly increasing with $c$, as shown in Fig.~\ref{fig:RelaxTimes}a, compatible with the constant structural heterogeneity length scale, shown in Fig.~\ref{fig:MultiScatt1}b. 
The short time scale $\tau_b$ also increases weakly with $c$, suggesting that the local motion is progressively hindered as the concentration rises.\\
\indent According to the proposed framework, this increasing hindrance should enhance the stability of the size-segregated domains. To guide the intuition, this effect can be compared to the increased depletion strength associated with a higher depletant concentration. 
Building on this analogy, we model the short-time diffusivity, $D_{fast}(\phi)$, with an Arrhenius law\cite{richteringComparisonViscosityDiffusion1995}: $D_{fast}(\phi)=D_0\exp\left[-E(\phi)/k_BT\right]$, where $D_0$ is the diffusion coefficient in the dilute sample, $c=0.1\%$, and $E(\phi)$ is a fit parameter quantifying the strength of the entropic effects in a sample with volume fraction $\phi$. 
The observed progressive slow-down in $D_{fast}$ with increasing concentration translates in $E$ increasing with $\phi$. In analogy with depletion interactions, where the interaction strength is linear in the depletant concentration, we assume that $E(\phi)=k\phi$ and fit our experimental data of $D_{fast}(\phi)$ to obtain $k\simeq7.3 \ k_BT$. We observe that the assumed linear dependence accounts fairly well for the observed decrease in diffusivity with increasing concentration, as shown in Fig.~\ref{fig:RelaxTimes}b, corroborating the hypothesis that the dynamic slow down is of entropic origin.\\
\indent This model for $E(\phi)$ helps us rationalize the observed formation of transient clusters between $c=1\%$ and $c=4\%$, as indicated by the emergence of a structural correlation peak in Fig.~\ref{fig:MultiScatt1}b and by the dynamic transition to coexisting slow and fast dynamics. 
We predict that this transition takes place as the strength of the entropic drive to size segregation, here quantified by $E(\phi)$, becomes stronger than thermal fluctuations, which tend to restore sample homogeneity.
Our fit result allows us to identify a threshold volume fraction, $\phi^*=k_BT/k\simeq 10\%$, corresponding to a surfactant concentration of $c=2.2\%$, in good agreement with our microscopy results.
In the range of volume fractions spanned by the concentrated regime, we find that $E$ varies between $2$ and $3k_B T$. Once again, exploiting the analogy with depletion interactions to guide our intuition, we observe that in this range of moderate volume fractions, the threshold for the percolation of clusters is set to higher interaction strengths, and that at such lower strengths, the equilibrium state of the system is a fluid of particle clusters\cite{lu_gelation_2008}. This is compatible with the experimental evidence that in our vesicle suspension, the structural heterogeneity remains limited to a few vesicle sizes, and that clusters are dynamic, exhibiting diffusive-like relaxation on time scales only slightly slower than that of individual vesicles.\\
\indent Although our physical interpretation of $E(\phi)$ as a proxy for the strength of entropic segregation is based on an analogy with depletion, the soundness of the predictions based on this quantity seems to suggest that a common theoretical framework might exist. 
%on the sample microstructure, SAXS and rheology show that even slightly below this threshold, down to $c=1\%$, the sample starts exhibiting features typical of concentrated systems, such as shear-thinning and large-scale heterogeneities. 
This motivates ongoing efforts towards a deeper understanding of the rheological and microstructural properties of these highly polydisperse, repulsive dispersions at intermediate volume fractions.

\section{Conclusions}
In this paper, we investigated the rheology, microscopic structure, and thermal dynamics of a polydisperse solution of soft vesicles across a wide range of volume fractions. Rheology displayed shear-thinning behavior even at relatively low concentrations, where it could not be explained by caging and glassy dynamics. This hinted at the presence of a structural relaxation time, much longer than the applied shear rates, the origin of which was unclear. Using a combination of X-ray and light scattering techniques to probe the microstructure, we demonstrated that the onset of shear-thinning corresponds to a well-defined structural transition. This is unexpected, given that independent measurements of interactions between individual vesicles of similar composition indicated purely repulsive forces.\\
\indent To explain this transition, we hypothesized that an effective attractive interaction of entropic origin may emerge as a result of crowding. 
This assumption was inspired by recent experimental findings on the structure and dynamics of large colloids in dense, bidisperse suspensions, indicating attractive interactions akin to depletion forces\cite{xuJammingDepletionExtremely2023}. Here, we formulated the assumption that a similar scenario may hold for highly polydisperse suspensions like ours.
This hypothesis enabled us to interpret the microscopic dynamics, measured by DDM. In the dilute regime, the dynamics are diffusive, but as the volume fraction exceeds $\phi^*\simeq 10\%$, glassy-like features appear. In particular, we found that the dynamics slow down and develop an intermediate plateau between two diffusive relaxation modes. We attributed this complex relaxation to the transient nature of the structural heterogeneities, and we quantified their $\phi$-dependent stability through an effective energy parameter. Our experimental data is compatible with an energy parameter increasing linearly with concentration, like in depletion forces. 
This allows us to predict the observed structural transition and shear-thinning behavior.\\
\indent Similar transitions are widespread in soft disordered materials and are typically attributed to attractive interactions between elementary constituents. In this work, we propose that, even in absence of such attractive interactions, these behaviors may emerge as the result of crowding and large polydispersity. Our findings suggest that the physics of dense and highly polydisperse suspensions still holds surprises and novel behaviors, highlighting the need for a deeper and more comprehensive understanding.

\section*{Author contributions}
Conceptualization: P.E., S.A., G.D., V.G. and R.P.
Software and Data curation: N.K., P.E., S.A.
Formal Analysis and Visualization: P.E., N.K. and S.A.
Investigation: N.K., P.E., S.A.
Methodology: P.E. and S.A.
Supervision and Project administration: P.E.
Founding acquisition: G.D.
Resources: V.G.
Writing - original draft: P.E., N.K, S.A., V.G. and R.P.
Validation: All authors.
Writing - review \& Edit: All authors.

\section*{Conflicts of interest}
There are no conflicts to declare. %In accordance with our policy on \href{https://www.rsc.org/journals-books-databases/journal-authors-reviewers/author-responsibilities/#code-of-conduct}{Conflicts of interest} please ensure that a conflicts of interest statement is included in your manuscript here.  Please note that this statement is required for all submitted manuscripts.  If no conflicts exist, please state that ``''.

\section*{Data availability}
Data are available upon request.
%A data availability statement (DAS) is required to be submitted alongside all articles. Please read our \href{https://www.rsc.org/journals-books-databases/author-and-reviewer-hub/authors-information/prepare-and-format/data-sharing/#dataavailabilitystatements}{full guidance on data availability statements} for more details and examples of suitable statements you can use.
 
\section*{Acknowledgments}
The authors acknowledge funding from the European Union’s Horizon 2020 research and innovation program under the Marie Skłodowska-Curie grant agreement No 955605, project YIELDGAP (https://yieldgap-itn.com). The authors are grateful to \textit{Procter \& Gamble} for kindly providing the vesicle dispersion samples. We acknowledge SOLEIL for provision of the synchrotron radiation facilities allocated to the GDR2019 CNRS/INRAE. We thank Javier Perez for assistance in using the beamline SWING. R.P. and S.A. acknowledge the support from the Erwin Schrödinger International Institute for Mathematics and Physics (ESI) in the frame of the thematic program {\it Linking Microscopic Processes to the Macroscopic Rheological Properties in Inert and Living Soft Materials}. 
R.P. acknowledges MUR-PRIN 2022ETXBEY, and MUR-PRIN 2022 PNRR P2022KA5ZZ,  funded by the European Union – Next Generation EU.

%%%REFERENCES%%%
\bibliography{rsc.bib} %You need to replace "rsc" on this line with the name of your .bib file

\providecommand*{\mcitethebibliography}{\thebibliography}
\csname @ifundefined\endcsname{endmcitethebibliography}
{\let\endmcitethebibliography\endthebibliography}{}
\begin{mcitethebibliography}{51}
\providecommand*{\natexlab}[1]{#1}
\providecommand*{\mciteSetBstSublistMode}[1]{}
\providecommand*{\mciteSetBstMaxWidthForm}[2]{}
\providecommand*{\mciteBstWouldAddEndPuncttrue}
  {\def\EndOfBibitem{\unskip.}}
\providecommand*{\mciteBstWouldAddEndPunctfalse}
  {\let\EndOfBibitem\relax}
\providecommand*{\mciteSetBstMidEndSepPunct}[3]{}
\providecommand*{\mciteSetBstSublistLabelBeginEnd}[3]{}
\providecommand*{\EndOfBibitem}{}
\mciteSetBstSublistMode{f}
\mciteSetBstMaxWidthForm{subitem}
{(\emph{\alph{mcitesubitemcount}})}
\mciteSetBstSublistLabelBeginEnd{\mcitemaxwidthsubitemform\space}
{\relax}{\relax}

\bibitem[Mewis and Wagner(2012)]{mewis2012colloidal}
J.~Mewis and N.~J. Wagner, \emph{Colloidal suspension rheology}, Cambridge university press Cambridge, 2012, vol.~10\relax
\mciteBstWouldAddEndPuncttrue
\mciteSetBstMidEndSepPunct{\mcitedefaultmidpunct}
{\mcitedefaultendpunct}{\mcitedefaultseppunct}\relax
\EndOfBibitem
\bibitem[Seifert(1997)]{seifert1997configurations}
U.~Seifert, \emph{Advances in physics}, 1997, \textbf{46}, 13--137\relax
\mciteBstWouldAddEndPuncttrue
\mciteSetBstMidEndSepPunct{\mcitedefaultmidpunct}
{\mcitedefaultendpunct}{\mcitedefaultseppunct}\relax
\EndOfBibitem
\bibitem[Guida(2010)]{guida2010thermodynamics}
V.~Guida, \emph{Advances in colloid and interface science}, 2010, \textbf{161}, 77--88\relax
\mciteBstWouldAddEndPuncttrue
\mciteSetBstMidEndSepPunct{\mcitedefaultmidpunct}
{\mcitedefaultendpunct}{\mcitedefaultseppunct}\relax
\EndOfBibitem
\bibitem[Muddle \emph{et~al.}(1983)Muddle, Higgins, Cummins, Staples, and Lyle]{muddle1983light}
A.~G. Muddle, J.~S. Higgins, P.~G. Cummins, E.~J. Staples and I.~G. Lyle, \emph{Faraday Discussions of the Chemical Society}, 1983, \textbf{76}, 77--92\relax
\mciteBstWouldAddEndPuncttrue
\mciteSetBstMidEndSepPunct{\mcitedefaultmidpunct}
{\mcitedefaultendpunct}{\mcitedefaultseppunct}\relax
\EndOfBibitem
\bibitem[Seth \emph{et~al.}(2014)Seth, Ramachandran, Murch, and Leal]{seth2014origins}
M.~Seth, A.~Ramachandran, B.~P. Murch and L.~G. Leal, \emph{Langmuir}, 2014, \textbf{30}, 10176--10187\relax
\mciteBstWouldAddEndPuncttrue
\mciteSetBstMidEndSepPunct{\mcitedefaultmidpunct}
{\mcitedefaultendpunct}{\mcitedefaultseppunct}\relax
\EndOfBibitem
\bibitem[Seth and Gary~Leal(2014)]{seth2014rheological}
M.~Seth and L.~Gary~Leal, \emph{Journal of Rheology}, 2014, \textbf{58}, 1619--1645\relax
\mciteBstWouldAddEndPuncttrue
\mciteSetBstMidEndSepPunct{\mcitedefaultmidpunct}
{\mcitedefaultendpunct}{\mcitedefaultseppunct}\relax
\EndOfBibitem
\bibitem[Porpora \emph{et~al.}(2020)Porpora, Rusciano, Guida, Greco, and Pastore]{porpora2020understanding}
G.~Porpora, F.~Rusciano, V.~Guida, F.~Greco and R.~Pastore, \emph{Journal of Physics: Condensed Matter}, 2020, \textbf{33}, 104001\relax
\mciteBstWouldAddEndPuncttrue
\mciteSetBstMidEndSepPunct{\mcitedefaultmidpunct}
{\mcitedefaultendpunct}{\mcitedefaultseppunct}\relax
\EndOfBibitem
\bibitem[Leivers \emph{et~al.}(2018)Leivers, Seddon, Declercq, Robles, and Luckham]{leivers2018measurement}
M.~Leivers, J.~M. Seddon, M.~Declercq, E.~Robles and P.~Luckham, \emph{Langmuir}, 2018, \textbf{35}, 729--738\relax
\mciteBstWouldAddEndPuncttrue
\mciteSetBstMidEndSepPunct{\mcitedefaultmidpunct}
{\mcitedefaultendpunct}{\mcitedefaultseppunct}\relax
\EndOfBibitem
\bibitem[Israelachvili(2011)]{israelachvili2011intermolecular}
J.~N. Israelachvili, \emph{Intermolecular and surface forces}, Academic press, 3rd edn, 2011\relax
\mciteBstWouldAddEndPuncttrue
\mciteSetBstMidEndSepPunct{\mcitedefaultmidpunct}
{\mcitedefaultendpunct}{\mcitedefaultseppunct}\relax
\EndOfBibitem
\bibitem[Chiruvolu \emph{et~al.}(1995)Chiruvolu, Israelachvili, Naranjo, Xu, Zasadzinski, Kaler, and Herrington]{chiruvolu1995measurement}
S.~Chiruvolu, J.~Israelachvili, E.~Naranjo, Z.~Xu, J.~Zasadzinski, E.~Kaler and K.~Herrington, \emph{Langmuir}, 1995, \textbf{11}, 4256--4266\relax
\mciteBstWouldAddEndPuncttrue
\mciteSetBstMidEndSepPunct{\mcitedefaultmidpunct}
{\mcitedefaultendpunct}{\mcitedefaultseppunct}\relax
\EndOfBibitem
\bibitem[Hunter and Weeks(2012)]{hunter2012physics}
G.~L. Hunter and E.~R. Weeks, \emph{Reports on progress in physics}, 2012, \textbf{75}, 066501\relax
\mciteBstWouldAddEndPuncttrue
\mciteSetBstMidEndSepPunct{\mcitedefaultmidpunct}
{\mcitedefaultendpunct}{\mcitedefaultseppunct}\relax
\EndOfBibitem
\bibitem[Verbeke(2012-2013)]{mastersthesis}
K.~Verbeke, \emph{MSc thesis}, KU Leuven, Kasteelpark Arenberg 1 bus 2200, B-3001 Heverlee, 2012-2013\relax
\mciteBstWouldAddEndPuncttrue
\mciteSetBstMidEndSepPunct{\mcitedefaultmidpunct}
{\mcitedefaultendpunct}{\mcitedefaultseppunct}\relax
\EndOfBibitem
\bibitem[Oikonomou \emph{et~al.}(2018)Oikonomou, Christov, Cristobal, Bourgaux, Heux, Boucenna, and Berret]{oikonomou2018design}
E.~Oikonomou, N.~Christov, G.~Cristobal, C.~Bourgaux, L.~Heux, I.~Boucenna and J.-F. Berret, \emph{Journal of colloid and interface science}, 2018, \textbf{525}, 206--215\relax
\mciteBstWouldAddEndPuncttrue
\mciteSetBstMidEndSepPunct{\mcitedefaultmidpunct}
{\mcitedefaultendpunct}{\mcitedefaultseppunct}\relax
\EndOfBibitem
\bibitem[Philippe \emph{et~al.}(2018)Philippe, Truzzolillo, Galvan-Myoshi, Dieudonn{\'e}-George, Trappe, Berthier, and Cipelletti]{philippe2018glass}
A.-M. Philippe, D.~Truzzolillo, J.~Galvan-Myoshi, P.~Dieudonn{\'e}-George, V.~Trappe, L.~Berthier and L.~Cipelletti, \emph{Physical Review E}, 2018, \textbf{97}, 040601\relax
\mciteBstWouldAddEndPuncttrue
\mciteSetBstMidEndSepPunct{\mcitedefaultmidpunct}
{\mcitedefaultendpunct}{\mcitedefaultseppunct}\relax
\EndOfBibitem
\bibitem[Mason and Weitz(1995)]{masonOpticalMeasurementsFrequencyDependent1995}
T.~G. Mason and D.~A. Weitz, \emph{Physical Review Letters}, 1995, \textbf{74}, 1250--1253\relax
\mciteBstWouldAddEndPuncttrue
\mciteSetBstMidEndSepPunct{\mcitedefaultmidpunct}
{\mcitedefaultendpunct}{\mcitedefaultseppunct}\relax
\EndOfBibitem
\bibitem[Mason \emph{et~al.}(1997)Mason, Ganesan, Van~Zanten, Wirtz, and Kuo]{masonParticleTrackingMicrorheology1997}
T.~G. Mason, K.~Ganesan, J.~H. Van~Zanten, D.~Wirtz and S.~C. Kuo, \emph{Physical Review Letters}, 1997, \textbf{79}, 3282--3285\relax
\mciteBstWouldAddEndPuncttrue
\mciteSetBstMidEndSepPunct{\mcitedefaultmidpunct}
{\mcitedefaultendpunct}{\mcitedefaultseppunct}\relax
\EndOfBibitem
\bibitem[Cicuta and Donald(2007)]{cicutaMicrorheologyReviewMethod2007}
P.~Cicuta and A.~M. Donald, \emph{Soft Matter}, 2007, \textbf{3}, 1449\relax
\mciteBstWouldAddEndPuncttrue
\mciteSetBstMidEndSepPunct{\mcitedefaultmidpunct}
{\mcitedefaultendpunct}{\mcitedefaultseppunct}\relax
\EndOfBibitem
\bibitem[Bayles \emph{et~al.}(2017)Bayles, Squires, and Helgeson]{baylesProbeMicrorheologyParticle2017}
A.~V. Bayles, T.~M. Squires and M.~E. Helgeson, \emph{Rheologica Acta}, 2017, \textbf{56}, 863--869\relax
\mciteBstWouldAddEndPuncttrue
\mciteSetBstMidEndSepPunct{\mcitedefaultmidpunct}
{\mcitedefaultendpunct}{\mcitedefaultseppunct}\relax
\EndOfBibitem
\bibitem[Edera \emph{et~al.}(2017)Edera, Bergamini, Trappe, Giavazzi, and Cerbino]{edera2017differential}
P.~Edera, D.~Bergamini, V.~Trappe, F.~Giavazzi and R.~Cerbino, \emph{Physical Review Materials}, 2017, \textbf{1}, 073804\relax
\mciteBstWouldAddEndPuncttrue
\mciteSetBstMidEndSepPunct{\mcitedefaultmidpunct}
{\mcitedefaultendpunct}{\mcitedefaultseppunct}\relax
\EndOfBibitem
\bibitem[Krall and Weitz(1998)]{krall_internal_1998}
A.~H. Krall and D.~A. Weitz, \emph{Physical Review Letters}, 1998, \textbf{80}, 778\relax
\mciteBstWouldAddEndPuncttrue
\mciteSetBstMidEndSepPunct{\mcitedefaultmidpunct}
{\mcitedefaultendpunct}{\mcitedefaultseppunct}\relax
\EndOfBibitem
\bibitem[Weitz \emph{et~al.}(1993)Weitz, Zhu, Durian, Gang, and Pine]{weitz1993diffusing}
D.~Weitz, J.~Zhu, D.~Durian, H.~Gang and D.~Pine, \emph{Physica Scripta}, 1993, \textbf{1993}, 610\relax
\mciteBstWouldAddEndPuncttrue
\mciteSetBstMidEndSepPunct{\mcitedefaultmidpunct}
{\mcitedefaultendpunct}{\mcitedefaultseppunct}\relax
\EndOfBibitem
\bibitem[Pastore \emph{et~al.}(2017)Pastore, Pesce, and Caggioni]{pastore2017differential}
R.~Pastore, G.~Pesce and M.~Caggioni, \emph{Scientific reports}, 2017, \textbf{7}, 43496\relax
\mciteBstWouldAddEndPuncttrue
\mciteSetBstMidEndSepPunct{\mcitedefaultmidpunct}
{\mcitedefaultendpunct}{\mcitedefaultseppunct}\relax
\EndOfBibitem
\bibitem[Cerbino and Trappe(2008)]{cerbino2008differential}
R.~Cerbino and V.~Trappe, \emph{Physical review letters}, 2008, \textbf{100}, 188102\relax
\mciteBstWouldAddEndPuncttrue
\mciteSetBstMidEndSepPunct{\mcitedefaultmidpunct}
{\mcitedefaultendpunct}{\mcitedefaultseppunct}\relax
\EndOfBibitem
\bibitem[Giavazzi \emph{et~al.}(2009)Giavazzi, Brogioli, Trappe, Bellini, and Cerbino]{giavazzi2009scattering}
F.~Giavazzi, D.~Brogioli, V.~Trappe, T.~Bellini and R.~Cerbino, \emph{Physical Review E}, 2009, \textbf{80}, 031403\relax
\mciteBstWouldAddEndPuncttrue
\mciteSetBstMidEndSepPunct{\mcitedefaultmidpunct}
{\mcitedefaultendpunct}{\mcitedefaultseppunct}\relax
\EndOfBibitem
\bibitem[Cavagna(2009)]{cavagna2009supercooled}
A.~Cavagna, \emph{Physics Reports}, 2009, \textbf{476}, 51--124\relax
\mciteBstWouldAddEndPuncttrue
\mciteSetBstMidEndSepPunct{\mcitedefaultmidpunct}
{\mcitedefaultendpunct}{\mcitedefaultseppunct}\relax
\EndOfBibitem
\bibitem[Janssen(2018)]{janssen2018mode}
L.~M. Janssen, \emph{Frontiers in Physics}, 2018, \textbf{6}, 97\relax
\mciteBstWouldAddEndPuncttrue
\mciteSetBstMidEndSepPunct{\mcitedefaultmidpunct}
{\mcitedefaultendpunct}{\mcitedefaultseppunct}\relax
\EndOfBibitem
\bibitem[Walde and Ichikawa(2001)]{walde2001enzymes}
P.~Walde and S.~Ichikawa, \emph{Biomolecular engineering}, 2001, \textbf{18}, 143--177\relax
\mciteBstWouldAddEndPuncttrue
\mciteSetBstMidEndSepPunct{\mcitedefaultmidpunct}
{\mcitedefaultendpunct}{\mcitedefaultseppunct}\relax
\EndOfBibitem
\bibitem[Murphy(2015)]{murphy2015fabric}
D.~S. Murphy, \emph{Journal of Surfactants and Detergents}, 2015, \textbf{18}, 199--204\relax
\mciteBstWouldAddEndPuncttrue
\mciteSetBstMidEndSepPunct{\mcitedefaultmidpunct}
{\mcitedefaultendpunct}{\mcitedefaultseppunct}\relax
\EndOfBibitem
\bibitem[Aime and Cipelletti(2019)]{aimeProbingShearinducedRearrangements2019a}
S.~Aime and L.~Cipelletti, \emph{Soft Matter}, 2019, \textbf{15}, 213--226\relax
\mciteBstWouldAddEndPuncttrue
\mciteSetBstMidEndSepPunct{\mcitedefaultmidpunct}
{\mcitedefaultendpunct}{\mcitedefaultseppunct}\relax
\EndOfBibitem
\bibitem[Cerbino \emph{et~al.}(2022)Cerbino, Giavazzi, and Helgeson]{cerbinoDifferentialDynamicMicroscopy2022}
R.~Cerbino, F.~Giavazzi and M.~E. Helgeson, 2022, \textbf{60}, 1079--1089\relax
\mciteBstWouldAddEndPuncttrue
\mciteSetBstMidEndSepPunct{\mcitedefaultmidpunct}
{\mcitedefaultendpunct}{\mcitedefaultseppunct}\relax
\EndOfBibitem
\bibitem[Nixon-Luke \emph{et~al.}(2022)Nixon-Luke, Arlt, K.~Poon, Bryant, and A.~Martinez]{nixon-lukeProbingDynamicsTurbid2022}
R.~Nixon-Luke, J.~Arlt, W.~C. K.~Poon, G.~Bryant and V.~A.~Martinez, \emph{Soft Matter}, 2022, \textbf{18}, 1858--1867\relax
\mciteBstWouldAddEndPuncttrue
\mciteSetBstMidEndSepPunct{\mcitedefaultmidpunct}
{\mcitedefaultendpunct}{\mcitedefaultseppunct}\relax
\EndOfBibitem
\bibitem[Xu and Mason(2023)]{xuJammingDepletionExtremely2023}
Y.~Xu and T.~G. Mason, \emph{Science Advances}, 2023, \textbf{9}, eadh3715\relax
\mciteBstWouldAddEndPuncttrue
\mciteSetBstMidEndSepPunct{\mcitedefaultmidpunct}
{\mcitedefaultendpunct}{\mcitedefaultseppunct}\relax
\EndOfBibitem
\bibitem[Corominas \emph{et~al.}(2011)Corominas, Quan, Yang, and Peers]{Corominas}
F.~Corominas, K.-M. Quan, Y.~Yang and K.~Peers, \emph{US Patent App. 12/984,663}, 2011\relax
\mciteBstWouldAddEndPuncttrue
\mciteSetBstMidEndSepPunct{\mcitedefaultmidpunct}
{\mcitedefaultendpunct}{\mcitedefaultseppunct}\relax
\EndOfBibitem
\bibitem[Seth \emph{et~al.}(2010)Seth, Ramachandran, and Leal]{seth2010dilution}
M.~Seth, A.~Ramachandran and L.~G. Leal, \emph{Langmuir}, 2010, \textbf{26}, 15169--15176\relax
\mciteBstWouldAddEndPuncttrue
\mciteSetBstMidEndSepPunct{\mcitedefaultmidpunct}
{\mcitedefaultendpunct}{\mcitedefaultseppunct}\relax
\EndOfBibitem
\bibitem[Giavazzi \emph{et~al.}(2017)Giavazzi, Edera, Lu, and Cerbino]{giavazzi2017image}
F.~Giavazzi, P.~Edera, P.~J. Lu and R.~Cerbino, \emph{The European Physical Journal E}, 2017, \textbf{40}, 1--9\relax
\mciteBstWouldAddEndPuncttrue
\mciteSetBstMidEndSepPunct{\mcitedefaultmidpunct}
{\mcitedefaultendpunct}{\mcitedefaultseppunct}\relax
\EndOfBibitem
\bibitem[ddm()]{ddmweb}
\emph{{{UCSB}} - {{DDMCalc License Terms}} | {{Helgeson Lab}}}, \url{https://helgeson.chemengr.ucsb.edu/ucsb-ddmcalc-license-terms}\relax
\mciteBstWouldAddEndPuncttrue
\mciteSetBstMidEndSepPunct{\mcitedefaultmidpunct}
{\mcitedefaultendpunct}{\mcitedefaultseppunct}\relax
\EndOfBibitem
\bibitem[Zemb and Lindner(2002)]{zemb_neutron_2002}
\emph{Neutron, {{X-rays}} and {{Light}}. {{Scattering Methods Applied}} to {{Soft Condensed Matter}}}, ed. T.~Zemb and P.~Lindner, Elsevier, 1st edn, 2002\relax
\mciteBstWouldAddEndPuncttrue
\mciteSetBstMidEndSepPunct{\mcitedefaultmidpunct}
{\mcitedefaultendpunct}{\mcitedefaultseppunct}\relax
\EndOfBibitem
\bibitem[Krieger and Dougherty(1959)]{krieger1959mechanism}
I.~M. Krieger and T.~J. Dougherty, \emph{Trans. Soc. Rheol}, 1959, \textbf{3}, 137--152\relax
\mciteBstWouldAddEndPuncttrue
\mciteSetBstMidEndSepPunct{\mcitedefaultmidpunct}
{\mcitedefaultendpunct}{\mcitedefaultseppunct}\relax
\EndOfBibitem
\bibitem[Cheng \emph{et~al.}(2011)Cheng, McCoy, Israelachvili, and Cohen]{cheng2011imaging}
X.~Cheng, J.~H. McCoy, J.~N. Israelachvili and I.~Cohen, \emph{Science}, 2011, \textbf{333}, 1276--1279\relax
\mciteBstWouldAddEndPuncttrue
\mciteSetBstMidEndSepPunct{\mcitedefaultmidpunct}
{\mcitedefaultendpunct}{\mcitedefaultseppunct}\relax
\EndOfBibitem
\bibitem[Xu \emph{et~al.}(2013)Xu, Rice, and Dinner]{xu2013relation}
X.~Xu, S.~A. Rice and A.~R. Dinner, \emph{Proceedings of the National Academy of Sciences}, 2013, \textbf{110}, 3771--3776\relax
\mciteBstWouldAddEndPuncttrue
\mciteSetBstMidEndSepPunct{\mcitedefaultmidpunct}
{\mcitedefaultendpunct}{\mcitedefaultseppunct}\relax
\EndOfBibitem
\bibitem[Saha \emph{et~al.}(2017)Saha, Chaudhuri, Godfrin, Mamak, Reeder, Hodgdon, Saveyn, Tripathi, and Bose]{saha2017impact}
A.~Saha, S.~Chaudhuri, M.~P. Godfrin, M.~Mamak, B.~Reeder, T.~Hodgdon, P.~Saveyn, A.~Tripathi and A.~Bose, \emph{Industrial \& Engineering Chemistry Research}, 2017, \textbf{56}, 899--906\relax
\mciteBstWouldAddEndPuncttrue
\mciteSetBstMidEndSepPunct{\mcitedefaultmidpunct}
{\mcitedefaultendpunct}{\mcitedefaultseppunct}\relax
\EndOfBibitem
\bibitem[Farr and Groot(2009)]{farrClosePackingDensity2009}
R.~S. Farr and R.~D. Groot, \emph{The Journal of Chemical Physics}, 2009, \textbf{131}, 244104\relax
\mciteBstWouldAddEndPuncttrue
\mciteSetBstMidEndSepPunct{\mcitedefaultmidpunct}
{\mcitedefaultendpunct}{\mcitedefaultseppunct}\relax
\EndOfBibitem
\bibitem[Percus and Yevick(1958)]{percusAnalysisClassicalStatistical1958}
J.~K. Percus and G.~J. Yevick, \emph{Physical Review}, 1958, \textbf{110}, 1--13\relax
\mciteBstWouldAddEndPuncttrue
\mciteSetBstMidEndSepPunct{\mcitedefaultmidpunct}
{\mcitedefaultendpunct}{\mcitedefaultseppunct}\relax
\EndOfBibitem
\bibitem[Mao \emph{et~al.}(1995)Mao, Cates, and Lekkerkerker]{maoDepletionForceColloidal1995}
Y.~Mao, M.~E. Cates and H.~N.~W. Lekkerkerker, \emph{Physica A: Statistical Mechanics and its Applications}, 1995, \textbf{222}, 10--24\relax
\mciteBstWouldAddEndPuncttrue
\mciteSetBstMidEndSepPunct{\mcitedefaultmidpunct}
{\mcitedefaultendpunct}{\mcitedefaultseppunct}\relax
\EndOfBibitem
\bibitem[Bibette \emph{et~al.}(1990)Bibette, Roux, and Nallet]{bibetteDepletionInteractionsFluidsolid1990}
J.~Bibette, D.~Roux and F.~Nallet, \emph{Physical Review Letters}, 1990, \textbf{65}, 2470--2473\relax
\mciteBstWouldAddEndPuncttrue
\mciteSetBstMidEndSepPunct{\mcitedefaultmidpunct}
{\mcitedefaultendpunct}{\mcitedefaultseppunct}\relax
\EndOfBibitem
\bibitem[Pednekar \emph{et~al.}(2018)Pednekar, Chun, and Morris]{pednekarBidispersePolydisperseSuspension2018}
S.~Pednekar, J.~Chun and J.~F. Morris, \emph{Journal of Rheology}, 2018, \textbf{62}, 513--526\relax
\mciteBstWouldAddEndPuncttrue
\mciteSetBstMidEndSepPunct{\mcitedefaultmidpunct}
{\mcitedefaultendpunct}{\mcitedefaultseppunct}\relax
\EndOfBibitem
\bibitem[Möbius \emph{et~al.}(2001)Möbius, Lauderdale, Nagel, and Jaeger]{mobiusSizeSeparationGranular2001}
M.~E. Möbius, B.~E. Lauderdale, S.~R. Nagel and H.~M. Jaeger, \emph{Nature}, 2001, \textbf{414}, 270--270\relax
\mciteBstWouldAddEndPuncttrue
\mciteSetBstMidEndSepPunct{\mcitedefaultmidpunct}
{\mcitedefaultendpunct}{\mcitedefaultseppunct}\relax
\EndOfBibitem
\bibitem[Pastore \emph{et~al.}(2022)Pastore, Giavazzi, Greco, and Cerbino]{pastore2022multiscale}
R.~Pastore, F.~Giavazzi, F.~Greco and R.~Cerbino, \emph{The Journal of Chemical Physics}, 2022, \textbf{156}, 164906--164917\relax
\mciteBstWouldAddEndPuncttrue
\mciteSetBstMidEndSepPunct{\mcitedefaultmidpunct}
{\mcitedefaultendpunct}{\mcitedefaultseppunct}\relax
\EndOfBibitem
\bibitem[Richtering and Mueller(1995)]{richteringComparisonViscosityDiffusion1995}
W.~Richtering and H.~Mueller, \emph{Langmuir}, 1995, \textbf{11}, 3699--3704\relax
\mciteBstWouldAddEndPuncttrue
\mciteSetBstMidEndSepPunct{\mcitedefaultmidpunct}
{\mcitedefaultendpunct}{\mcitedefaultseppunct}\relax
\EndOfBibitem
\bibitem[Lu \emph{et~al.}(2008)Lu, Zaccarelli, Ciulla, Schofield, Sciortino, and Weitz]{lu_gelation_2008}
P.~J. Lu, E.~Zaccarelli, F.~Ciulla, A.~B. Schofield, F.~Sciortino and D.~A. Weitz, \emph{Nature}, 2008, \textbf{453}, 499--503\relax
\mciteBstWouldAddEndPuncttrue
\mciteSetBstMidEndSepPunct{\mcitedefaultmidpunct}
{\mcitedefaultendpunct}{\mcitedefaultseppunct}\relax
\EndOfBibitem
\bibitem[Anderson \emph{et~al.}(2010)Anderson, Donaldson, Zeng, and Israelachvili]{andersonDirectMeasurementDoubleLayer2010}
T.~H. Anderson, S.~H. Donaldson, H.~Zeng and J.~N. Israelachvili, \emph{Langmuir}, 2010, \textbf{26}, 14458--14465\relax
\mciteBstWouldAddEndPuncttrue
\mciteSetBstMidEndSepPunct{\mcitedefaultmidpunct}
{\mcitedefaultendpunct}{\mcitedefaultseppunct}\relax
\EndOfBibitem
\end{mcitethebibliography}
\bibliographystyle{rsc.bst} %the RSC's .bst file

\section*{Appendix}
\subsection*{Conversion from surfactant concentration to volume fraction}
\begin{figure}[h!]
    \centering
    \includegraphics[width=1.0\linewidth]{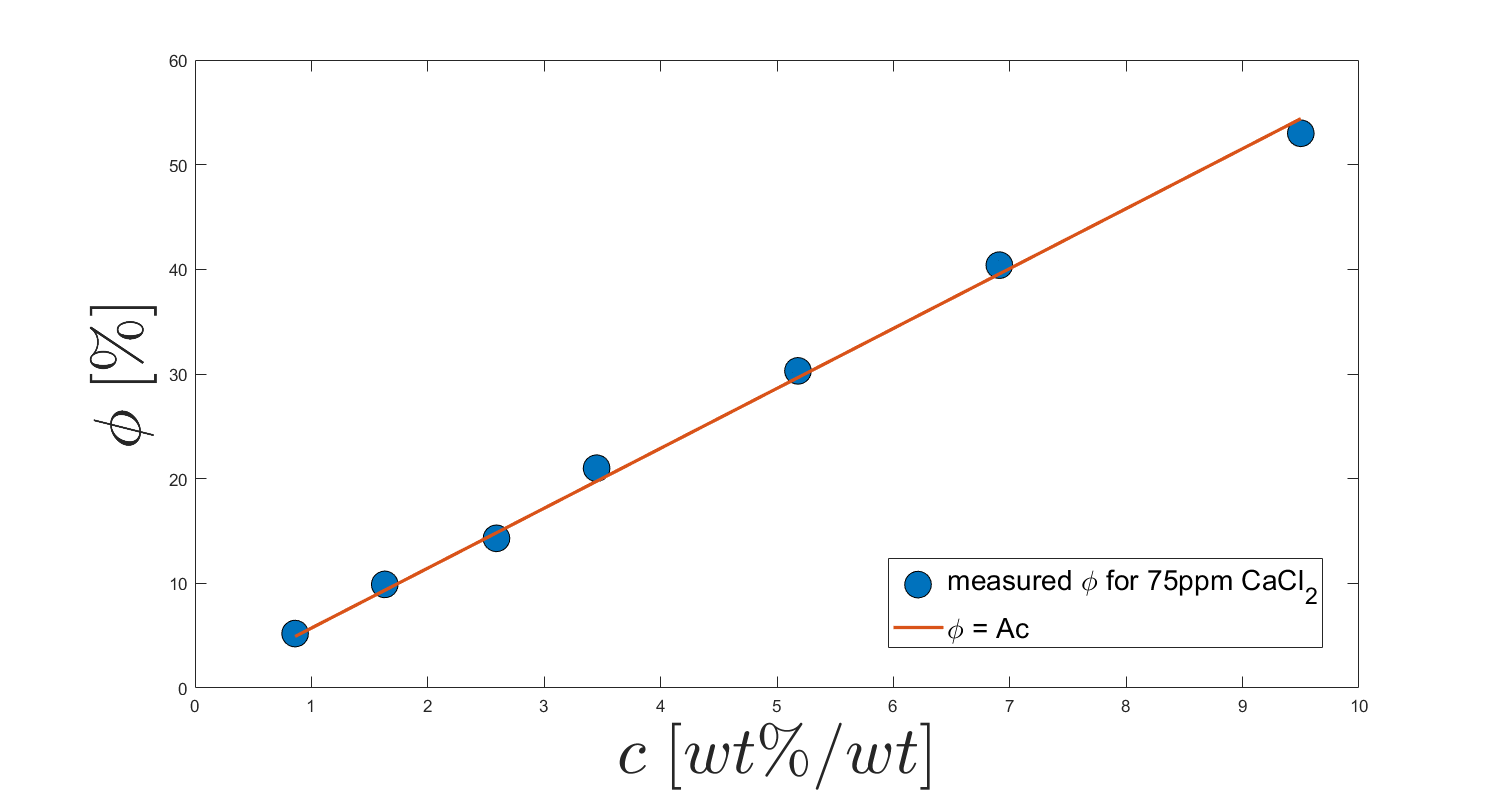}
    \caption{Correspondence between vesicles volume fraction and surfactant concentration for a white base with $75ppm$ calcium chloride. Blue circles: measured volume fraction values for each $c$. Red line: linear fitting ($\phi=5.73 c$) to these data.}
    \label{fig:vf_conc}
\end{figure}

The volume fraction of the white bases is measured following the method described by Seth et al. \cite{seth2010dilution}. Fig.~\ref{fig:vf_conc} shows the measured volume fraction values for a white base containing $75ppm$ calcium chloride at various surfactant concentrations~$c$. These data are well-captured by a linear model, $\phi=A c$, with $A=5.73$, as shown by the red line in Fig.~\ref{fig:vf_conc}. For our samples, which contain $450ppm$ $CaCl_2$, the volume fraction at $c=13\%$ is measured to be $\phi=58\%$. We then assume that $c$ and $\phi$ are related by a similar linear relationship, with a different prefactor, $A^\prime=4.46$. The $\phi$ values obtained applying this scaling for all surfactant concentrations are provided in Tab.~\ref{tbl:vf-conc}.
\begin{table}[h!]
\small
\centering
  \caption{\ volume fraction values}
  \label{tbl:vf-conc}
  \begin{tabular*}{0.4\textwidth}{@{\extracolsep{\fill}}cc}
    \hline
    concentration [$\%$] & volume fraction [$\%$] \\
    \hline
    0.01   & 0.044  \\
    0.1   & 0.44  \\
    0.3  & 1.3 \\
    1  & 4.5 \\
    2  & 9 \\
    4  & 18 \\
    6  & 27 \\
    8  & 36 \\
    10  & 45 \\
    11  & 49 \\
    12  & 54 \\
    13  & 58 \\
    \hline
  \end{tabular*}
\end{table}
\subsection*{Interaction potential}
In agreement with a previous work in the literature\cite{leivers2018measurement}, we model the repulsion between vesicles in terms of an electrostatic double-layer interaction potential, which we express as a function of the surface-to-surface distance, $D$, in units of the thermal energy $k_BT$, as reported by Israelachvili\cite{israelachvili2011intermolecular}: %eq. 14.51
\begin{equation}
    \frac{W_{EDL}(D)}{k_BT} =  \frac{64\pi\rho_\infty \gamma^2R}{\kappa^2} e^{-\kappa D} \,  ,
    \label{eq.edl}
\end{equation}
\noindent where: 
\begin{itemize}
    \item $\rho_\infty$ is the bulk electrolyte ion concentration. Here, 450ppm of $CaCl_2$ correspond to $\rho_\infty=4 mMol/l=2.4\cdot 10^{24} m^{-3}$;
    \item $\gamma=\tanh(ze\psi_0/4k_BT)$, with $z$ the electrolyte valence, $e$ the elementary charge, $\psi_0$ the surface potential. As a realistic estimate, we take $\psi_0=70mV$\cite{leivers2018measurement,andersonDirectMeasurementDoubleLayer2010}
    \item $\kappa^{-1}$ is the Debye length. It corresponds to:
\begin{equation}
    \frac{1}{\kappa} = \sqrt{\frac{\varepsilon_0\varepsilon k_BT}{2 \rho_\infty e^2}} \,  ,
\end{equation}
    with $\varepsilon_0\varepsilon$ the dielectric permittivity of the fluid medium. For water at room temperature, $\varepsilon \approx 80$ and $\varepsilon_0\varepsilon k_BT/2e^2=5.5\cdot 10^7 m^{-1}$. For 1:2 electrolytes such as $CaCl_2$, one obtains $\kappa^{-1}=0.176/\sqrt{[CaCl_2]} nm$, where $[CaCl_2]$ is the concentration in $Mol/l$. For us, $[CaCl_2]=4\cdot 10^{-3}$, thus $\kappa^{-1}=2.8~nm$.
    \item $R$ is the vesicle radius. For this estimation, we use the median radius, $R=140 nm$
\end{itemize}
\begin{figure*}[t]
    \centering
    \includegraphics[width=0.95\linewidth]{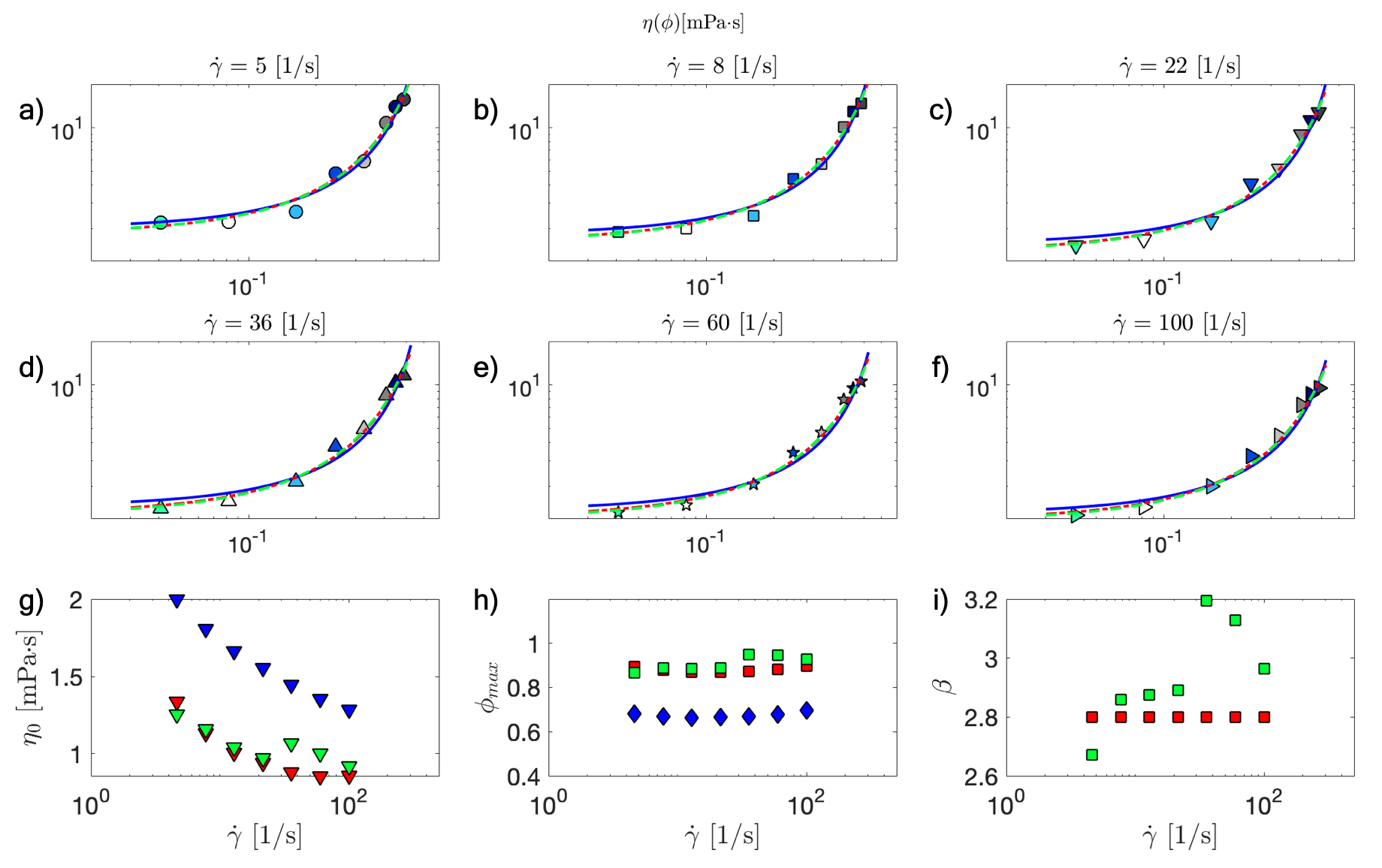}
    \caption{(a-f) Concentration dependent viscosity at different rates. The divergence of the viscosity can be described by different rheological models: Krieger-Dougherty model (blue), Mode Coupling Theory with three free parameters (green), or with two free parameters (red).
    (g) Rate dependent $\eta_0$: low concentration limit viscosity. (h) Maximum volume fraction. (i) MCT Viscosity divergence experiment.}
    \label{fig:RheoModels}
\end{figure*}
With the numbers above, we get an exponentially-decaying interaction energy with prefactor $W_0=W_{EDL}(0)=200 k_BT$, representing the extrapolated interaction energy at surface contact.

We define the interaction distance, $d^*$, as the distance at which $W_{EDL}(d^*)=k_BT$. From Eq.~\ref{eq.edl}, we obtain: $d^*=\kappa^{-1}\ln(W_0/k_BT)=15~nm$.
\subsection*{Rheology and comparison of fitting models}
Fig.~\ref{fig:RheoModels} shows the comparison between different rheological models: Krieger-Dougherty equation, used in the main text
\cite{krieger1959mechanism}, 
\begin{equation}
    \eta=\eta_s\cdot (1- \phi/\phi_{max})^{-2.5\cdot \phi_{max}} \  ,
\end{equation}
Mode Coupling Theory (MCT) model
\begin{equation}
    \eta=\eta_s\cdot (\phi_{max}-\phi)^{-\beta} \,
    \label{eq.MCT}
\end{equation}
with two (red) or three (green) fit parameters. When left as a free fit parameter, the exponent $\beta$ is relatively noisy (green symbols in Fig.~\ref{fig:RheoModels}i).
The value of $\beta$ can be fixed to $2.8$  (red symbols), without decreasing the quality of the fits.
Both models closely reproduce the rheological behavior. 
The maximum volume fraction is stable with the shear rate for both models, and is higher than the prediction for hard and weakly polydisperse spheres.
KD predicts a steeper dependence, with a smaller $\phi_{max}$: $\phi_{max}^{KD}\simeq 0.7<\phi_{max}^{MCT}\simeq0.9 $.\\
Both models provide a rate-dependent low-concentration-viscosity $\eta_0$, with the rate-dependence being an indirect sign of shear-thinning.

\subsection*{Small Angle X-ray Scattering and background subtraction}

Differences in the capillaries thickness reflect in differences in the baseline, due to glass and solvent (water) scattering.
\begin{figure}[h!]
    \centering
    \includegraphics[width=0.95\linewidth]{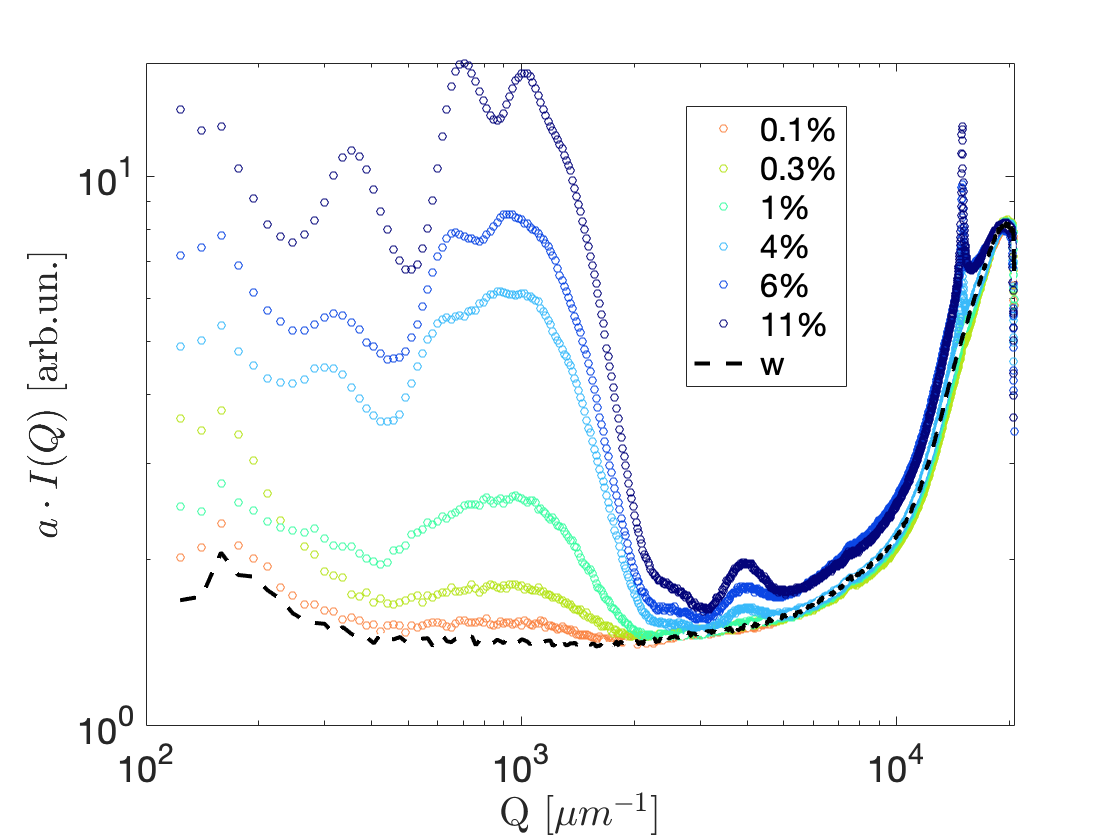}
    \caption{X-ray signal for the high $Q$ sensor before background subtraction. 
        $I_B(Q)$ background signal, scattered by a capillary filled with water and 450 ppm CaCO3 (black and white dashed line).
        Rescaled signal $a\cdot I_R(Q)$.}
    \label{fig:XraysBack}
\end{figure}
To properly remove the background (black and white dashed line in Fig.~\ref{fig:XraysBack}), we define a sample dependent constant, $a$, such that for each sample, $a\cdot I(Q)$ superimposes to the water signal at the higher $Q$s: $Q\simeq2\cdot 10^4 \mu m ^{-1}$. The signal in the main text (Fig.~\ref{fig:MultiScatt1}a) is obtained as $I_R(Q)-\frac{I_B(Q)}{a}$.

\subsection*{Multiple scattering}
\begin{figure}[h!]
\centering
  \includegraphics[width=1.0 \linewidth]{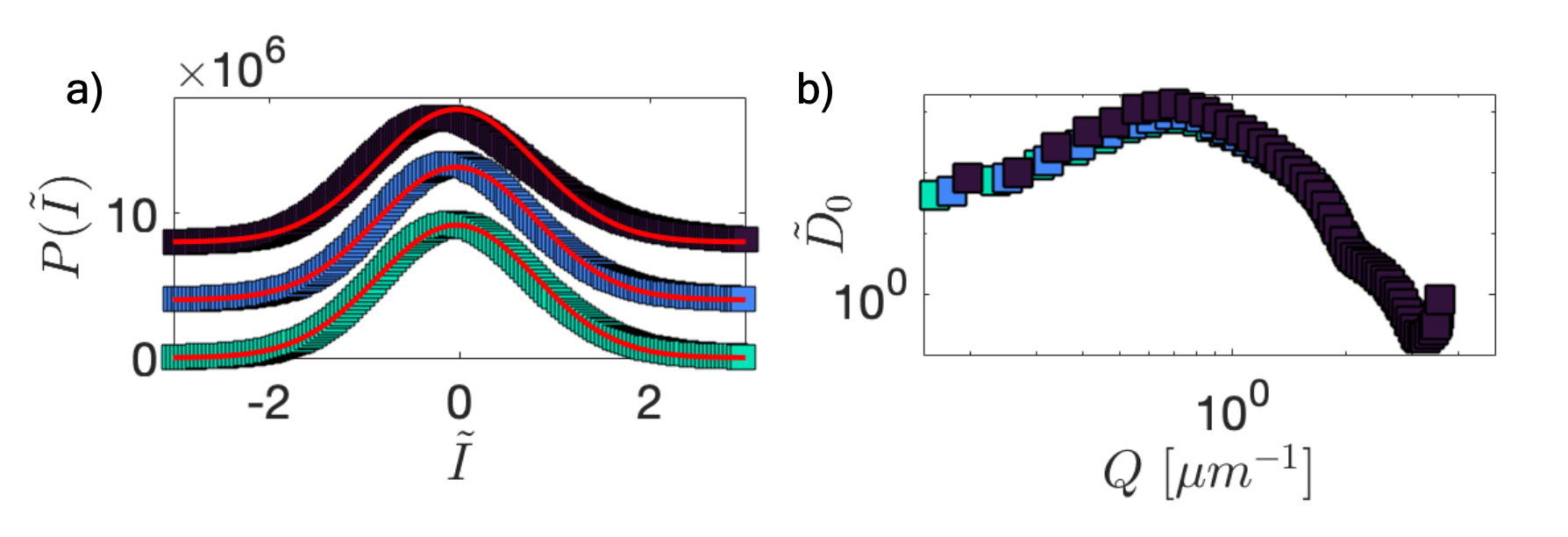}
  \caption{Microscopy, static indicators. 
  a) Symbols: normalized intensity histograms for $c=4\%$ (cyan), 6\% (blue), 11\% (black). Data have been vertically shifted for clarity. Red lines: Gaussian fits to the distributions. 
  b) Rescaled image power spectrum $\Tilde{D}_0(Q;c)$ for the same concentrations.
  % \SA{suggestion: move title for y axis to the left, rearrange panel order according to the order they are mentioned in the text: c-a-b}
  }
  \label{fig:MultiScatt2}
  \end{figure}

\paragraph*{Intensity histograms}
The influence of multiple scattering on DDM measurements can be evaluated by analyzing the histograms of image intensities, $P(\tilde{I})$, obtained for different sample concentrations (Fig.~\ref{fig:MultiScatt2}a). The raw intensity histogram, $P(I)$, is transformed into $P(\tilde{I})$ using the standardized variable: $\tilde{I}=(I-\mu)/\sigma $, where $\mu$ is the mean intensity, used to horizontally translate the histograms, and $\sigma$ is the standard deviation, which normalizes and scales the data. This transformation aligns the histograms from images acquired at different sample concentrations, facilitating their comparison. After alignment, the histograms can be analyzed for deviations from a Gaussian distribution, which may indicate the presence of multiple scattering effects that alter the statistical properties of the measured scattering intensities.
In our samples, we observe such deviations for $c=11\%$, while up to $c=6\%$, $P(\tilde{I})$ is well-described by a Gaussian distribution.\\
\paragraph*{Rescaled image power spectrum}
The rescaled image power spectrum, $\Tilde{D}_0(Q;c)$, provides a means to identify the wavevector range where multiple scattering effects may arise.
It is defined as $\Tilde{D}_0(Q;c)=a(c)\cdot D_0(Q)$, where $D_0(Q)$ represents the static power spectrum, and $\alpha(c)$ is a concentration-dependent scaling factor. The factor $\alpha(c)$ is defined, so that the curves superimpose in the high-$Q$ regime, where single scattering dominates. 
In the single scattering regime, changes in concentration simply scale the intensity without altering the power spectrum, as in this regime the scattering process is linear.
Any deviations from the expected superposition in $D_0(Q;c)$ indicate departure from linearity, potentially due to multiple scattering effects. In our analysis, such deviation is observed only for $c=11\%$ at lower $Q$ (Fig.~\ref{fig:MultiScatt2}b). For this sample, we restrict our analysis to $Q>0.9\ \mu m^{-1}$, while for the rest of the dispersions, the whole accessible $Q$ range is considered.

\subsection*{Vesicle dispersions DDM analysis}

\begin{figure*}[t]
\centering
    \includegraphics[width=0.9 \linewidth]{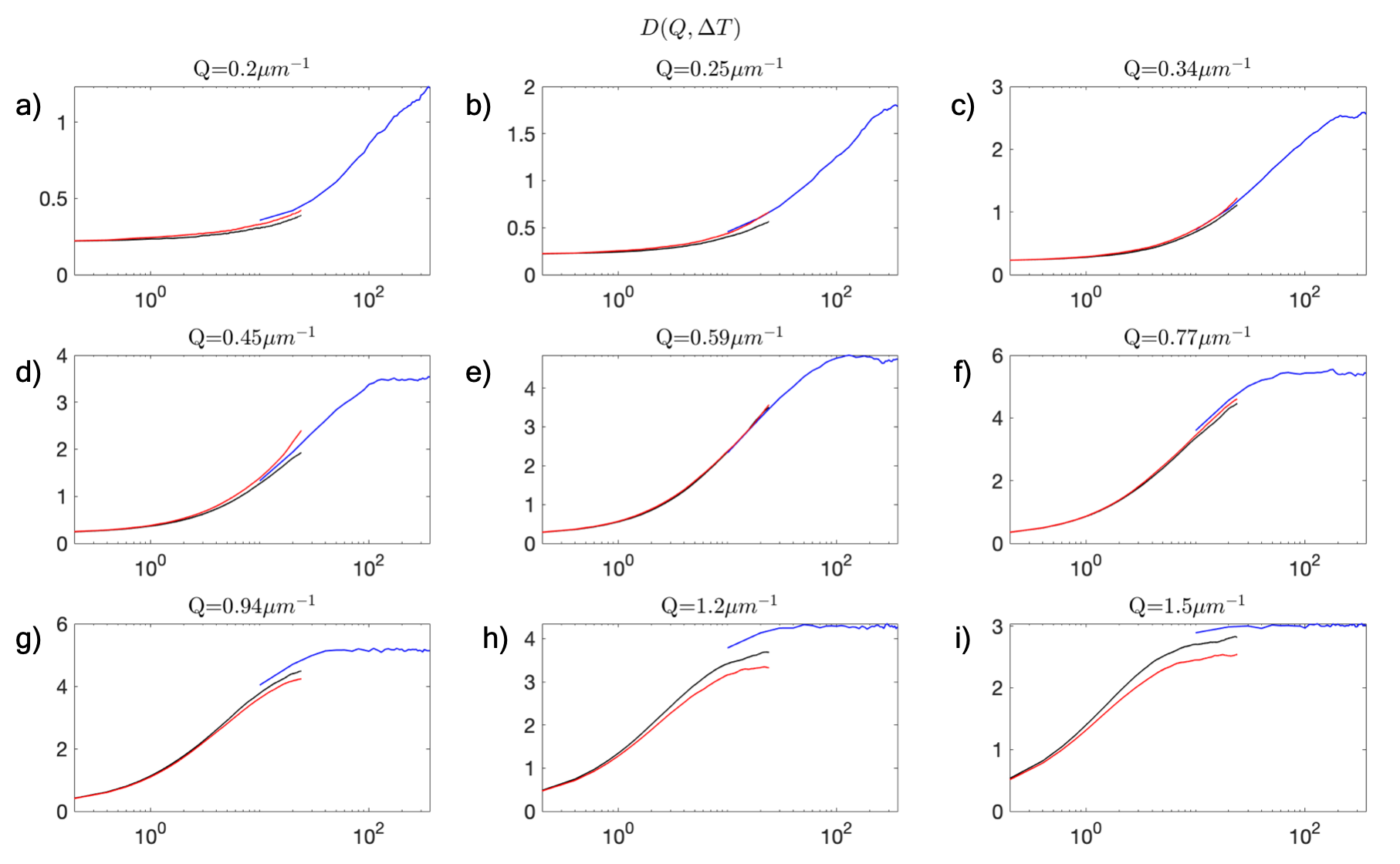}
    \caption{Non-normalized structure functions at different rates.
    Experiments at different rates has a little discrepancy.
    The discrepancy is not due to systematic temporal evolution of the properties.
    We merge the functions that do not reach complete decorrelation within the duration of shorter experiments.
    }
    \label{fig:DDM_Merging}
\end{figure*}
\begin{figure}[t!]
    \centering
    \includegraphics[width=0.8\linewidth]{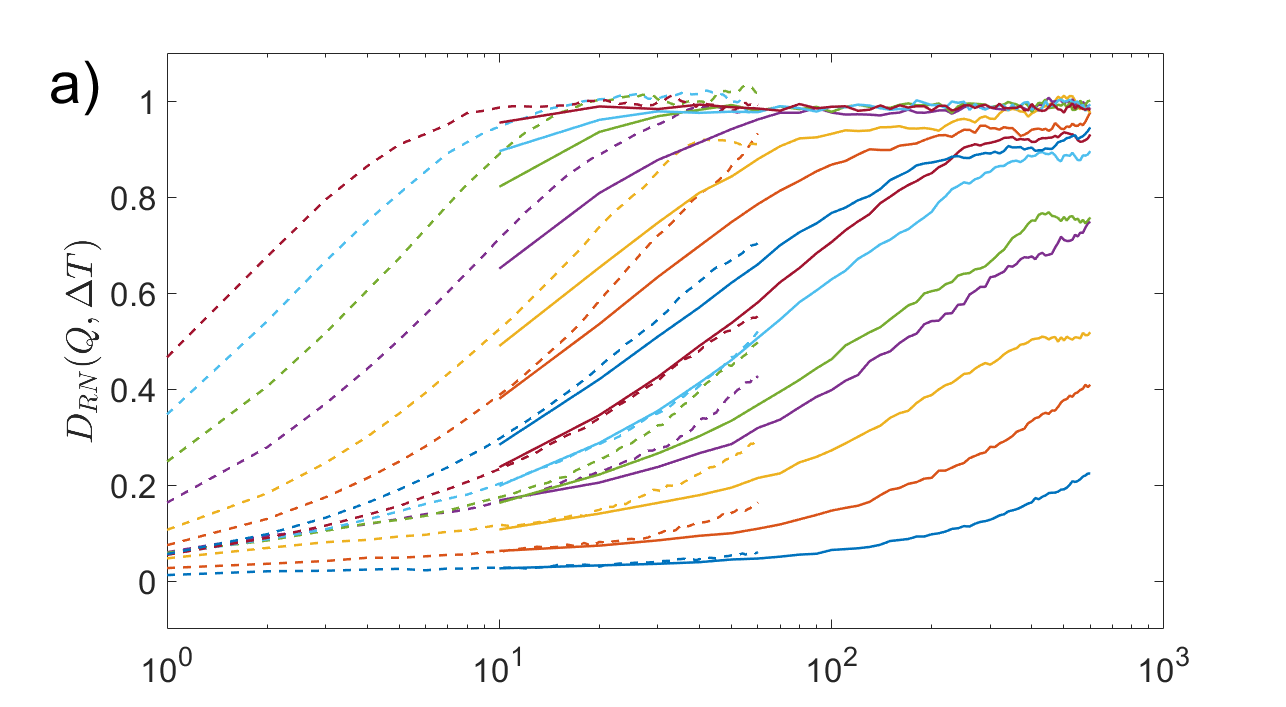}
    \includegraphics[width=0.8\linewidth]{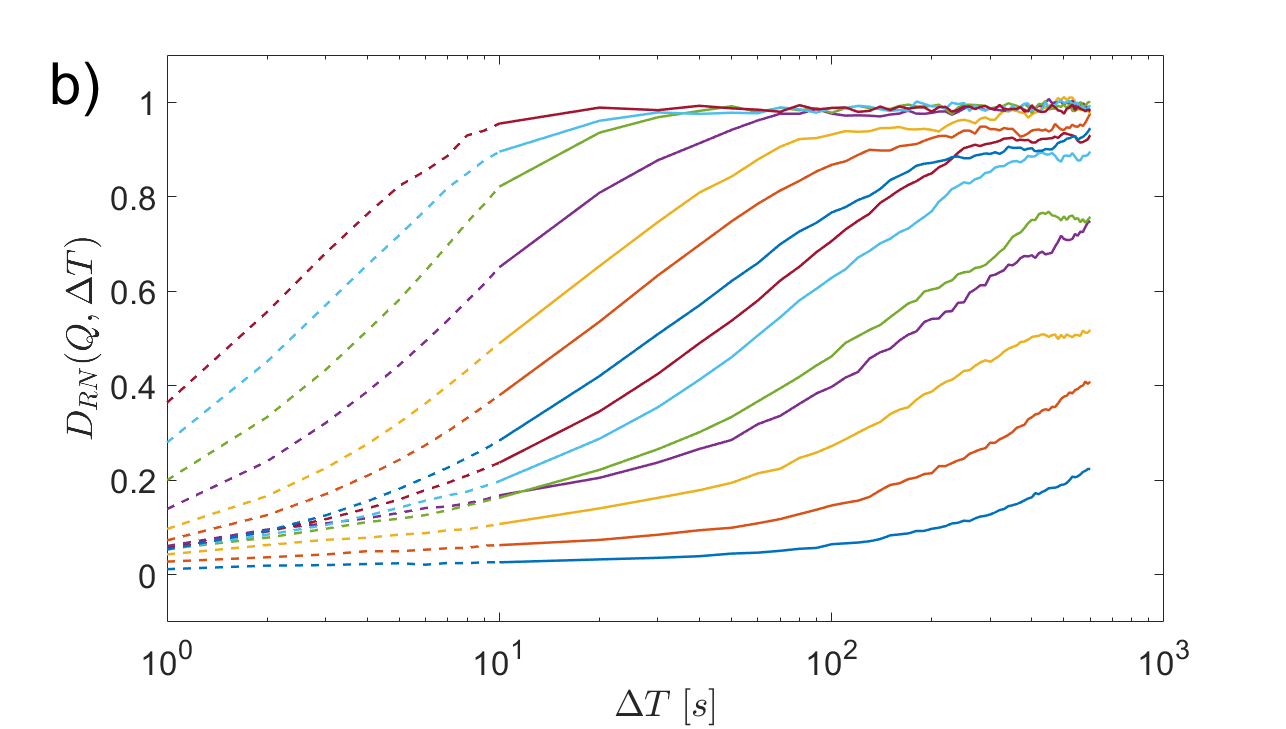}
    \caption{Renormalized image structure function before (a) and after (b) the merging process of the two acquisitions. Solid and dashed lines represent the $0.1$ and $1Hz$ frame rates data, respectively.}
    \label{Merging}
\end{figure}

The DDM analysis of the vesicle dispersions depends on the investigated concentration. 
For the dilute samples of concentrations $c=0.1$ and $1\%$, we obtain two image series at $0.1Hz$ ($180$ images) and $5Hz$ ($900$ images). 
For each frame rate, we compute the image structure function, $D(Q,\Delta T)$. 
Experiments at different rates do not perfectly match, but the difference is small and does not present any trivial systematics (Fig.~\ref{fig:DDM_Merging}). We thus merge the two experiments at different rates through a multiplicative rescaling of the structure function derived from the faster measurement.\\
\indent To obtain the intermediate scattering function, we fit $D(Q,\Delta T)$ with a single stretched exponential decay function, $D(Q,\Delta T)=~A(Q)(1-\exp\{-(\Delta T/\tau(Q))^{0.825}\}+B(Q)$, 
where $A(Q)$ and $B(Q)$ are the two $Q$-dependent parameters accounting, respectively, for the particles signal and for the noise. This allows us to obtain the ISF (Fig.~\ref{fig:Dilut}a) as:
\begin{equation}
f(Q,\Delta T)=\frac{D(Q,\Delta T)-B(Q))}{A(Q)}\,
\end{equation}
\indent For the concentrated samples of $c=4, 6$, and $11\%$, the suspension dynamics are recorded at two distinct acquisition rates of $0.1$ and $1Hz$, each consisting of $120$ images. For each frame rate, we calculate again the dynamic ($D(Q,\Delta T)$) and the static ($D_0(Q)$) power spectra without any background subtraction. For such concentrated samples, the signal is dominated by the particles, and hence no correction is needed. 
The normalized dynamic power spectrum is computed using $\frac{D(Q,\Delta T)-B}{D_0(Q)-B}$, with the noise value $B$ selected to achieve proper normalization across the whole $Q$ range.\\ 
\indent The normalized functions from the two acquisitions must be combined into a single curve to access both small and large timescales.
The normalized power spectra exhibit a small discrepancy, shown in Fig.~\ref{Merging}a.
This deviation is likely due to a subtle sample aging effect that leads to a slow down of the dynamics.
To address these discrepancies, 
% in normalized power spectra between the $1Hz$ and $0.1Hz$ acquisition rates, particularly at larger wavevectors ((Fig.~\ref{Merging} (a))), 
we apply a renormalization on the logarithm of the intermediate scattering function of the $1Hz$ dataset.
First, we compute the function $h(Q,\Delta T)=-ln(f(Q,\Delta T))$ for both datasets.
We take the ratio $\kappa(Q)=h_{1Hz}(Q)/h_{0.1Hz}(Q)$ at $T=10s$, and we use $\kappa (Q)$ to renormalize $h$ before merging them: $h^R_{1Hz}(Q,T)=\kappa(Q)\cdot h_{1Hz}(Q,T)$.
Once we have merged the two
\begin{equation}
  h^M(Q,T)=\begin{cases}
    h^R_{1Hz}(Q,T) & \text{if  $T<10$}\\
    h_{0.1Hz}(Q,T) & \text{if $T\geq 10$},
  \end{cases}
\end{equation}
we obtain the renormalized intermediate scattering function as: 
\begin{equation}
    f=\exp(-h^M)
\end{equation}
(Fig.~\ref{Merging}b). After merging the curves, the ISFs are averaged over bins of similar wavevector values, yielding a total of 12 curves. Each of these curves is then fitted to a single or a double stretched exponential decay, as described in the main text.

\newpage
\subsection*{DDM intermediate scattering functions}

\begin{figure}[t!]
\centering
    \includegraphics[width=1.0 \linewidth]{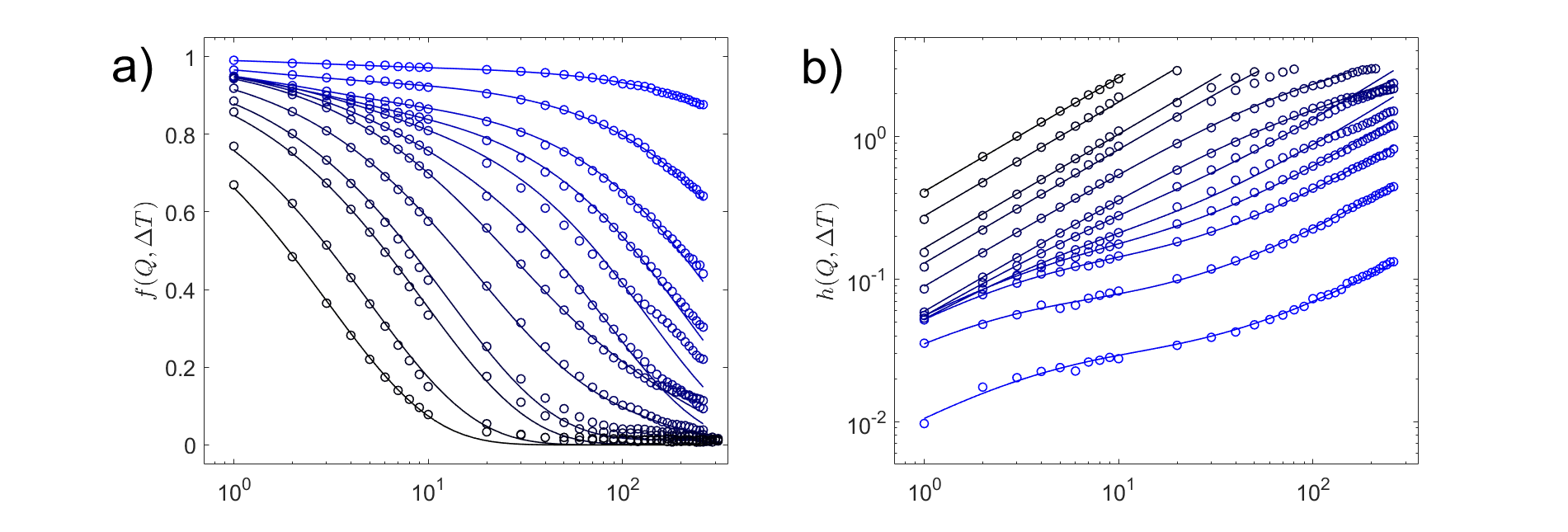}
    \includegraphics[width=1.0 \linewidth]{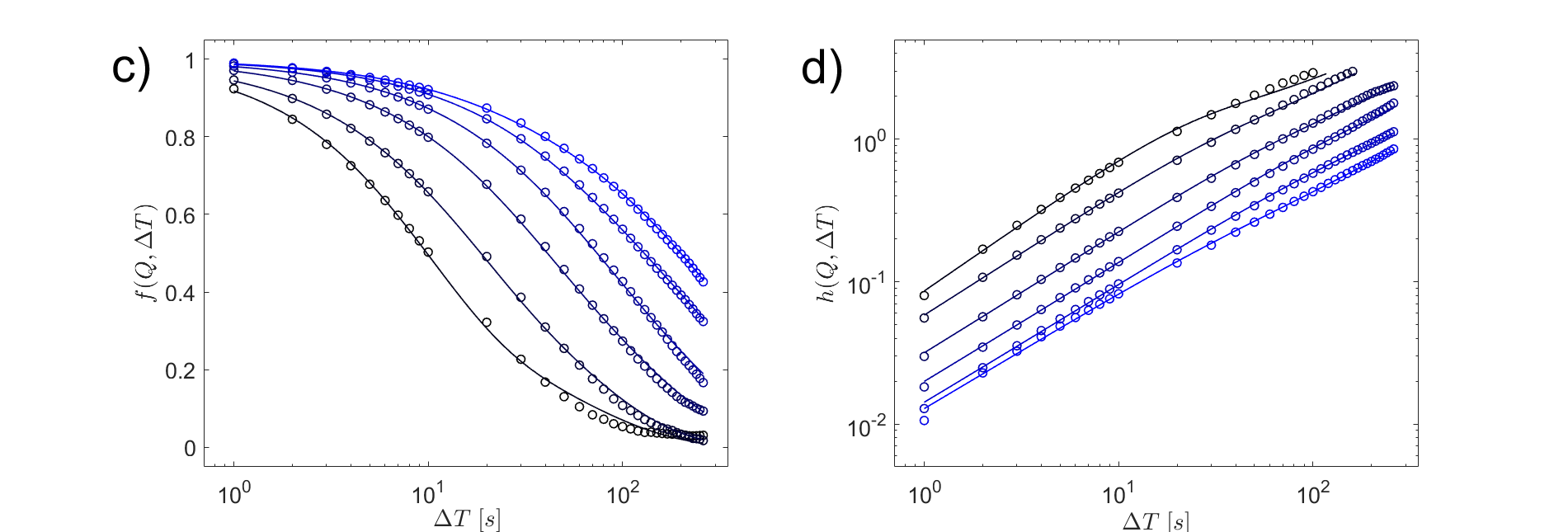}
    \caption{Intermediate scattering functions $f(Q,\Delta T)$, and $h(Q,\Delta T)=-\ln[f(Q,\Delta T)]$  for $c=6\%$ (a-b) and $c=11\%$ (c-d). The range of accessible wave vectors for $6\%$ is the same as for $c=4\%$.
    For $c=11\%$, the accessible range of wave vectors is limited to $Q>1.8\ \mu m^{-1}$, because of multiple scattering. %(see SI for further discussion). 
    The same two-processes picture discussed in Fig.~\ref{fig:Dyn_04} holds for the two samples. The color code is the same as of Fig.~\ref{fig:Dyn_04}, and in particular, the highest $Q$ shown for $6\%$ corresponds to the highest $Q$ presented for $11\%$ (in black).}
    \label{fig:IFSconcentr}
\end{figure}

\begin{figure}[ht!]
\centering
  \includegraphics[width=0.95\linewidth]{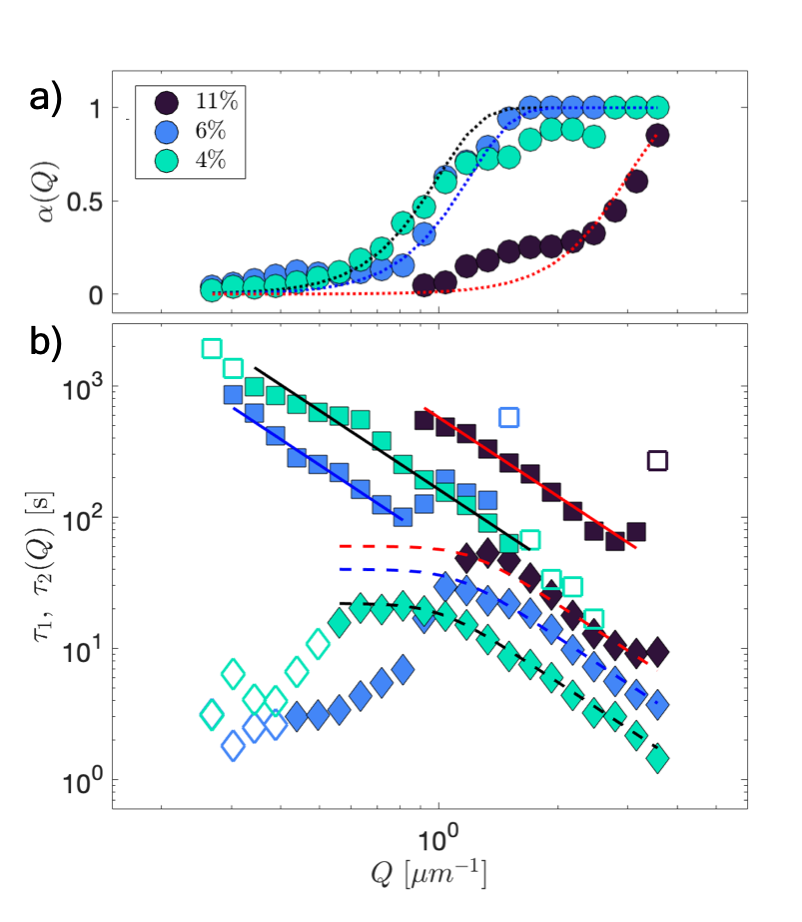}
  \caption{Microscopic dynamics fit parameters. a) Amplitude of the faster relaxation process $\alpha(Q)$ for $c=4\%$ (green), $6\%$ (blue)  and $11\%$ (black). b) Fast (diamonds) and slow (squares) relaxation times. All data are fitted with the models introduced in Fig.~\ref{fig:Dyn_04}.
}
\label{fig:Concentr}
\end{figure}

Fig.~\ref{fig:IFSconcentr} illustrates the intermediate scattering functions, as well as the function $h$ of the samples with $c=6, 11\%$.
Tab.~\ref{tbl:stretch} presents the stretching exponents obtained from  fitting the ISFs of samples with $c\geq4\%$ to Eq.~\ref{eq:isf_double}, while in Fig.~\ref{fig:Concentr} we plot the parameters derived from this fit.
%\newline   

\begin{table}[ht!]
\small
  \caption{\ stretching exponents}
  \label{tbl:stretch}
  \begin{tabular*}{0.48\textwidth}{@{\extracolsep{\fill}}cccc}
    \hline
    concentration [$\%$] & volume fraction [$\%$] & $\beta_1$ & $\beta_2$\\
    \hline
    4   & 18 & 0.8  & 0.9  \\
    6   & 27 & 0.8  & 0.9 \\
    11  & 49 & 1    & 0.7 \\
    \hline
  \end{tabular*}
\end{table}

\newpage
\subsection*{DLS intermediate scattering functions}
To be completed...

\begin{figure}[ht!]
\centering
    \includegraphics[width=.8 \linewidth]{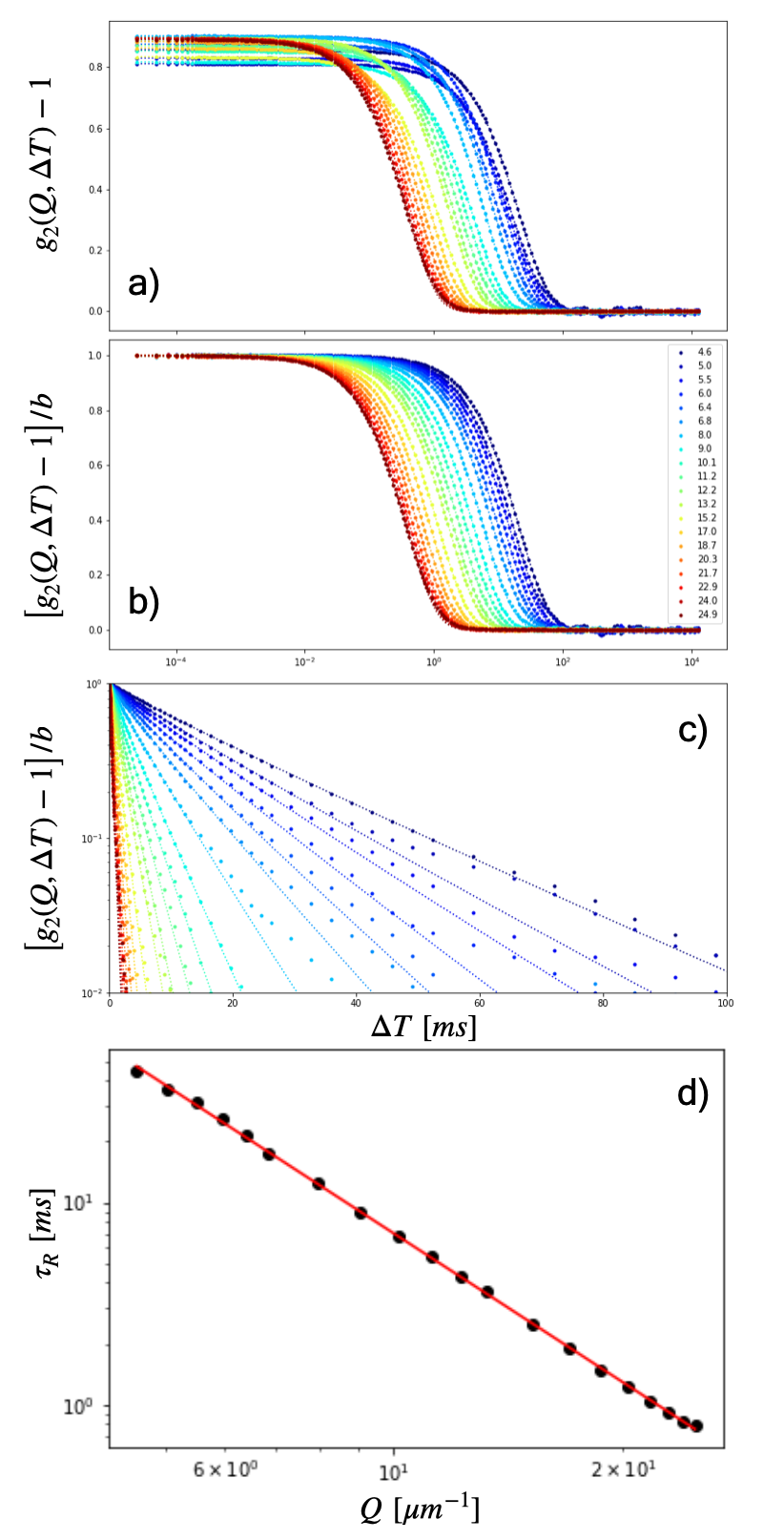}
    \caption{
    a) Homodine DLS Intermediate scattering functions, $g_2-1$, for the sample in dilute conditions $c=0.1$. Wavevectors go from $4.6\ \mu m^{-1}$ (blue) to $24.9\ \mu m^{-1}$ (red), as indicate in the legend.
    b) Normalized  $\frac{g_2-1}{b}$ in lin-log scale.
    c) Normalized  $\frac{g_2(Q,\Delta T)-1}{b}$ in log-lin scale.
    The data are fitted with stretched exponential decays, with stretching exponent $\beta_{DLS}=0.94$.$\exp(-(\Delta T/\tau)^{\beta_{DLS}})$
    d) Average relaxation time $\tau_R(Q)$, fitted with $AQ^{-p}$, resulting in $A=4.25$, $p=2.45$. 
    Because of the difference between DLS $g_2$ and DDM $f$, that are measured respectively in homodine and heterodine, we compare the DDM relaxation time $\tau^DDM$ with $2\cdot \tau_R^{DLS}$ Fig.~\ref{fig:Dilut}b.
    }
    \label{fig:SI_DLS}
\end{figure}

%%%END OF MAIN TEXT%%%


% \balance


\end{document}